\documentclass[usegraphicx]{mn2e}

\begin{document}

\def\Model#1#2{{\tt #1}{\tt #2}}
\newfont{\normal}{rptmr scaled 20000}
\title[Numerical simulations of interacting gas-rich barred galaxies]
      {Numerical simulations of interacting gas-rich barred galaxies.
       Vertical impact of small companions.}
\author[I.~Berentzen et al.]
       {I.~Berentzen$^{1,2}$\thanks{Email: iberent@uni-sw.gwdg.de},
        E.~Athanassoula$^2$,
        C.H.~Heller$^3$, 
        and K.J.~Fricke$^1$\\
        $^1$Universit\"ats--Sternwarte, Geismarlandstra\ss e 11,
            D-37083 G\"ottingen, Germany\\
        $^2$Observatoire de Marseille, 2 Place Le Verrier,
            F-13248 Marseille Cedex 4, France \\
        $^3$Georgia Southern University, Department of Physics,
            Statesboro, GA 30460, U.S.A. \\  }
\maketitle
\begin{abstract}
 We investigate the dynamical effects of an interaction between an
 {\em initially} barred galaxy and a small spherical companion
 using an $N$-body/SPH algorithm. In the models described here the
 small companion passes through the disc of the larger galaxy
 near-perpendicular to its plane. The impact positions and times
 are varied with respect to the phase of the bar and the dynamical
 evolution of the disc.

 The interactions produce expanding ring structures, offset bars,
 spokes, and other asymmetries in the stars and gas. These
 characteristic signatures of the interaction are present in the disc
 for about 1\,Gyr.  We find that in some cases it is possible to
 destroy the bar while keeping the disc structure. In general
 the central impacts cause larger damage to the bar and the disc
 than the peripheral ones. The interaction tends to accelerate the
 transition from a strongly barred galaxy to a weakly or non-barred
 galaxy.
 The final disc morphology is determined more by the impact
 position relative to the bar rather than the impact time.
\end{abstract}
\section{Introduction}
 The evolution of disc galaxies is driven through both internal and
 external processes. Internal instabilities in the disc often give
 rise to the formation of a bar, which as many as half of all disc
 galaxies are now known to harbour. In the last few years
 near-infrared (NIR) observations (Pompea \& Rieke 1990; Rix 1993;
 Rix \& Rieke 1993; Seigar \& James 1998; Eskridge et al. 2000) have
 confirmed that bars are a very common feature in disc galaxies and
 by no means an exception.
 The presence of a bar changes both the kinematics and mass (stellar
 and gas) distributions within the disc and can give raise to
 dynamical resonances (e.g., Sellwood \& Wilkinson 1993). Likewise,
 an only recently recognised important external agent is the
 interaction with small companions. A statistical survey by
 Zaritsky et al. (1993, 1997) gave a lower limit to the average
 number of companions for field spiral galaxies of $1.2\pm0.3$.
 Since these two processes are thought to be so common, it is a
 reasonable assumption that the interaction between a barred disc
 galaxy and a small companion would be a common event.

 Although only a small fraction of the mass of a galaxy, gas can
 significantly influence the evolution of disc galaxies (e.g.,
 Shlosman \& Noguchi 1993). Due to the torques of a stellar bar,
 the dissipative gas can be driven towards the galactic centre and
 channelled toward the inner kpc (Shlosman, Begelman \& Frank 1990;
 Athanassoula 1992b) driving active galactic nuclei (AGN) or nuclear
 starbursts \cite{hs94} and creating nuclear and circumnuclear discs
 and rings.  Recent NIR-Observations \cite{pkp} have shown a
 relation between the circumnuclear star formation in barred
 galaxies and the circumnuclear spiral or ring structure present in
 these galaxies.  The shape of dust lanes inside the bar, coinciding
 with the location of shocks in the gasflow, reveals information
 about the internal dynamics of the bar and disc \cite{ath92b}. Many
 numerical simulations including a dissipative disc component have
 emphasised the importance of the gas in isolated barred galaxies
 (e.g., Shlosman \& Noguchi 1993; Friedli \& Benz 1993; Heller
 \& Shlosman 1994; Berentzen et al. 1998) and interacting galaxies
 as well (e.g., Noguchi 1987; Hernquist \& Weil 1993; Barnes \&
 Hernquist 1996).

 Since the pioneering work of Toomre \& Toomre \shortcite{tt72} much
 effort has been spent to investigate galaxy-galaxy interactions
 (e.g., see the review by Barnes \& Hernquist 1992, and references
 therein). Noguchi \shortcite{nog88} started with numerical
 simulations to study the formation of bars in tidal interactions
 and the response of the gas to a tidal bar. Gerin, Combes
 \& Athanassoula \shortcite{gca} have performed both 2-D and 3-D
 $N$-body simulations to study the response of an initial bar formed
 in a globally unstable disc \cite{op73} to a tidal encounter with a
 spherical perturber. They studied the effect of a close encounter,
 both in and out of plane, with trajectories with pericenters
 outside the disc. This work has been continued by Sundin \&
 Sundelius \shortcite{ss} and Sundin, Donner \& Sundelius
 \shortcite{sds}, who, using 2D $N$-body simulations, studied the
 response of a bar, which had been {\em induced} by a satellite on a
 planar parabolic orbit. In these works, the change in pattern speed,
 angular momentum distribution and resonances in the disc were
 investigated for different orbits and masses of the perturber.
 Further work, concentrating on the formation of collisionally
 induced rings, has been performed by Athanassoula, Puerari \& Bosma
 (1997, hereafter APB97). These purely collisionless
 simulations have shown that off-centred impacts of a sufficiently
 massive companion hitting the inner parts of a barred disc galaxy
 can displace the bar to one side, causing asymmetries and the
 formation of rings. We should note that off-centred bars may also
 form spontaneously in galaxies.

 In this paper we focus upon the evolution of a gas-rich barred disc
 galaxy which is perturbed by the impact of a less massive spherical
 companion galaxy. An $N$-body/SPH algorithm is used to evolve the
 stellar and gas components of the two systems.  In the models
 described here the small companion passes through the disc of the
 larger galaxy near-perpendicular to its plane. The impact position
 and time are varied with respect to the bar and to the evolutionary
 phase of the isolated barred galaxy.  The mass of the companion
 galaxy has been chosen such that the interaction can be considered
 as a perturbation to the bar and disc. The set of encounter
 parameters has been deduced from the previous simulations by APB97
 and chosen as to produce the strongest effects, like asymmetries,
 off-centred bars and rings. We describe how the evolution of the
 bar strength, pattern speed, and gas inflow rate are affected by
 the interaction.


\section{Methods}

 The method consists of an $N$-body algorithm to evolve the
 collisionless component, representing the stars and dark matter,
 combined with a smoothed particle hydrodynamics (SPH) algorithm to
 evolve the dissipative component, representing the gas (e.g.,
 review by Monaghan 1992). For the simulations presented in this
 paper we use the hybrid $N$-body/SPH code which is described in
 detail by  Heller \shortcite{hel91} and Heller \& Shlosman
 \shortcite{hs94}. The algorithm employs such features as a
 spatially varying smoothing length, a hierarchy of time bins to
 approximate individual particle timesteps, a viscosity ``switch''
 to reduce the effects of viscous shear, and the special purpose
 GRAPE-3AF hardware to compute the gravitational forces and
 neighbour interaction lists (Sugimoto et al. 1990; Steinmetz 1996).
 Besides speed in the direct force summation, the GRAPE hardware has
 the additional advantage that it does not impose any constraints
 on the spatial distribution of the particles.

 The model of the disc galaxy (or {\em host} galaxy) is composed of
 a stellar and gaseous disc, embedded in a hot spherical dark matter
 halo. An isothermal equation of state is used for the gas
 component. The spherical companion galaxy (or {\em perturber})
 consists of stars only.

\subsection{Initial conditions}

\subsubsection{Host galaxy and companion}

 The model parameters for the host and the companion galaxy are
 selected similar to those used by APB97, so as to allow a
 meaningful comparison between the collisionless and dissipative
 models.  The host galaxy is a barred model referred to as {\bf mb}
 in APB97. Our model here, however, differs from {\bf mb} in that
 about 23\% of the mass of the stellar disc has been replaced by a
 gaseous disc. The radial scale-length and total mass of the disc
 have been retained in the model.

 Both the stellar and the gaseous disc are initially setup with a
 Kuzmin-Toomre (hereafter KT) projected radial surface density
 profile (Kuzmin 1956; Toomre 1963)
\begin{equation}
 \Sigma_{\rm KT}(R) = \frac{1}{2 \pi}\, {a\, M_{\rm KT}}
             \cdot \left( R^2 + a^2 \right)^{-3/2}.
\end{equation}
 In the above $R$, $M_{\rm KT}$ and $a$ are the cylindrical radius, the
 total mass and the radial scale-length, respectively. The resulting
 truncated mass profile for the disc is then set up with the disc 
 mass $M_{\rm D}\!=\!M_{\rm s}+M_{\rm g}$, where $M_{\rm s}$ and
 $M_{\rm g}$ are the masses of the stellar and the gaseous disc,
 respectively, which have been truncated at a radius $R_{\rm D}$
 The vertical disc profile follows the $\mbox{{\rm sech}}^2(z/z_i)$
 distribution of an isothermal sheet \cite{spi42}, where $z_i$ is 
 the vertical scale-height with $i\!=\!{\rm s}$ and
 $i\!=\!{\rm g}$ for the stellar and gaseous disc, respectively.

 Both the halo and the companion have a generic Plummer (hereafter Pl)
 density profile \cite{plu11}
\begin{equation}
 \rho_{\rm{Pl}}(r) = \frac{3\,b^2\,M_{\rm Pl}}{4 \pi}
           ( r^2 + b^2 )^{-5/2} ,
\end{equation}
 where $r$ is the spherical radius. $M_{\rm Pl}$ and $b$ denote
 the total mass and radial scale-length for each individual 
 component. The truncated mass profile for the halo and the 
 companion are set up with a cut-off mass $M_j$ and radial
 scale-length $b_j$ with $j\!=\!{\rm H}$ and ${\rm C}$,
 respectively, with the cut-off radius $r_j$.

 The initially spherical halo, not being in virial equilibrium with
 the embedded disc, is allowed to relax in the potential of the
 dynamically {\em frozen} disc, whose mass is slowly turned-on over
 time following the spline function
 \begin{equation}
  u(t) = \left\{
       \begin{array} {l@{\hspace{0.8cm} \mbox{for} \quad}l}
         \left( 3 - 2 \frac{t}{t_0} \right) \cdot
         \left(\frac{t}{t_0}\right)^2 & t \leq t_0 \\
         1.0 & t > t_0 \\
         \end{array} \right. ,
 \end{equation}
 where the turn-on time in our units has been chosen as
 $t_0\!=\!60$. The halo is relaxed for a total time period of
 $\Delta t\!=\!120$.
 From the resulting potential we then assign velocities to the
 disc particles, with the radial dispersion based on Toomre's
 stability criterion \cite{too64} and correcting for asymmetric
 drift.

\begin{figure}
 \begin{center}
 \includegraphics[angle=-90, scale=0.44]{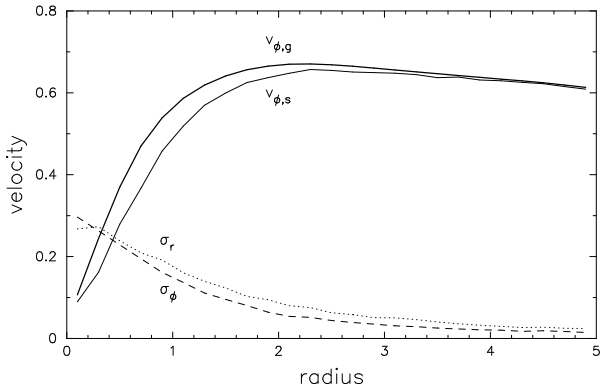}
 \end{center}
 \caption{Initial velocity distribution of the host model \Model{I}{0}.
  We show the mean tangential velocity of both the stellar (thin full line) and
  the gaseous disc (thick full line) as a function of radius. The radial and
  tangential velocity disperion of the stellar disc is shown with dotted and
  dashed lines, respectively.}
 \label{fig01}
\end{figure}

\begin{table}
\caption{Initial model parameters}
\label{init}
\begin{tabular}{lccrlccc} 
   & {\bf Type} &  ${\bf N}$  & {\bf M}$_{\rm {\bf D}}$ &
 {\bf a} & {\bf R}$_{\rm {\bf D}}$ & {\bf z$_{\bf s/g}$}\\[1.2ex]
 {\bf Disc}  &    &         &      &        &      &          \\
 {\hspace{1ex} -- stars} & KT & 13\,500 & 0.54 &  1.0   &  5.0 &  0.20 \\
 {\hspace{1ex} -- gas}   & KT & 10\,000 & 0.16 &  1.0   &  5.0 &  0.05 \\[3ex]
   & {\bf Type} &  ${\bf N}$  & {\bf M}$_{\rm {\bf H}}$ & {\bf b}$_{\rm {\bf
 H}}$ & {\bf r}$_{\rm {\bf H}}$ & \\[1.2ex]
 {\bf Halo}            &    &         &      &        &      &          \\
 {\hspace{1ex} -- stars} & Pl & 32\,500 & 1.30 &  5.0   & 10.0 &  \\[3ex]
   & {\bf Type} &  ${\bf N}$  & {\bf M}$_{\rm {\bf C}}$ & {\bf b}$_{\rm {\bf
 C}}$ & {\bf r}$_{\rm {\bf C}}$ & \\[1.2ex]
  {\bf Comp.}   &    &         &      &        &      &     \\
 {\hspace{1ex} -- stars} & Pl & 10\,000 & 0.40 &  0.195 &  3.0 &  \\
\end{tabular}
\end{table}

 The spherical companion galaxy is similar to the model referred to
 as {\bf csd} in APB97. The mass $M_{\rm C}$ of the companion has
 been chosen such that the mass ratios of companion to the host disc
 and to total host mass $M_{\rm tot}\!=\!M_{\rm D}+M_{\rm H}$ are,
 respectively, $M_{\rm C}/M_{\rm D}\!=\!0.57$ and
 $M_{\rm C}/M_{\rm tot}\!=\!0.20$. Based on the results of APB97, 
 we expect to obtain with these mass ratios the formation of spokes
 and rings following the interaction. The initial model parameters
 are summarised in Table~\ref{init}.
 The initial velocity distribution of the disc is shown in Fig.~\ref{fig01}.
 The tangential velocity curve has its turn-over radius at roughly 2a
 and stays flat in the outer disc.

\subsubsection{Units}
 
 The adopted units for mass, distance, and time are
 $M\!= 6\cdot10^{10}$\,M$_\odot$, $\rm{R}\!=\!3$\,kpc and
 $\tau\!=\!10^7$ yr, respectively, for which the gravitational
 constant G is unity. The dynamical time is
 $\tau_{\rm dyn} \!\equiv\! ( R^3_{\rm h}/{\rm G}\,M_{\rm h})^{1/2} \!=\!
 4.8\cdot10^7$ yr, where $M_{\rm h}$ is the total mass within a
 sphere of radius equal to the half mass radius $r_{\rm h}$, which
 is, after relaxation of the halo, approximately 8.5\,kpc or in
 terms of the disc scale-length, $2.8\,{\rm a}$. The initial stellar
 disc rotation period in these units then corresponds to
 $t_{\rm rot}\!\equiv\!2\pi\tau_{\rm dyn} \approx 3\!\cdot\!
 10^8 yr$.  A fixed gravitational softening length of
 $\epsilon\!=\!0.375$\,kpc is used for all particles. An isothermal
 equation of state is used for the gas with a sound speed of
 $v_{\rm s}\!=\!12$\,km s$^{-1}$.  The corresponding thermal
 temperature of the gas is 10$^4$\,K.

\subsubsection{Interaction parameters}

 Before adding the companion to the model we evolve the isolated
 barred host galaxy for $\Delta t \!=\! 300$, or $3 \cdot 10^9$ yr. 
 The initial centre of mass positions and velocities of the host and
 companion have then been obtained by integrating their orbits
 backward in time, starting from the impact time $t_{\rm imp}$ at
 which the centre of mass of the companion lies in the disc plane.
 During this integration the disc (with bar) and halo particles are
 frozen with respect to each other, but allowed to rotate as a
 single system with the ({\em negative}) angular frequency or
 pattern speed $\Omega_{\rm p}(t)$ of the bar, as determined from
 the isolated disc model (hereafter \Model{I}{0}). The companion is
 represented by a point mass for this integration.  This
 approximation is sufficient since the encounter is fast enough that
 dynamical friction does not significantly modify the orbit.

\begin{table}
\caption{Model and interaction description}
\label{init2}
\begin{tabular}{lclll} 
 {\bf Model}       & {\bf Name}   & {\bf $t_{\rm imp}$} & {\bf Bar}
    & {\bf $t_{\rm end}$} \\[2ex] 
 Isolated model    & \Model{I}{0} & $\cdots$    &            & $t\!=\!300$
 \\[0.5ex]
 Central impact    & \Model{C}{1} & $t\!=\  60$  & strong & $t\!=\!210$ \\
                   & \Model{C}{2} & $t\!=\!150$ & weak  & $t\!=\!300$
 \\[0.5ex]
 Major axis impact & \Model{A}{1} & $t\!=\ 60$  & strong  & $t\!=\!210$ \\
                   & \Model{A}{2} & $t\!=\!150$ & weak    & $t\!=\!300$
 \\[0.5ex]
 Minor axis impact & \Model{B}{1} & $t\!=\ 60$  & strong  & $t\!=\!210$ \\
                   & \Model{B}{2} & $t\!=\!150$ & weak   & $t\!=\!300$ 
\end{tabular}
\end{table}

 The impact velocity $v_{\rm imp}$ of the companion has been chosen
 to be of the order of $4\,v_{\rm esc}$, with $v_{\rm esc}$ being
 the escape velocity from the centre of the host system. The
 hyperbolic orbit is then integrated backward for a time period of
 $\Delta t\!=\!30$, which places the companion well outside the halo
 of the disc system. The simulations presented in this paper will be
 confined to almost perpendicular passages. Impacts along
 trajectories 30 degrees (and sometimes higher) from the disc normal
 still yield ring shaped features \cite{lyt76}. The effects of
 oblique impacts have been studied by Toomre \shortcite{too78},
 Athanassoula \shortcite{ath99} and APB97.
 An overview of the isolated and interaction models is given in
 Table~\ref{init2}. The columns, from left to right, give the type
 and the name of the model, the bar impact time $t_{\rm imp}$, the
 strength of the bar at $t_{\rm imp}$ and the end time $t_{\rm end}$
 of each simulation.

\section{Results}

\subsection{Model \Model{I}{0} : Isolated barred galaxy}

\begin{figure*}
 \includegraphics[scale=0.65]{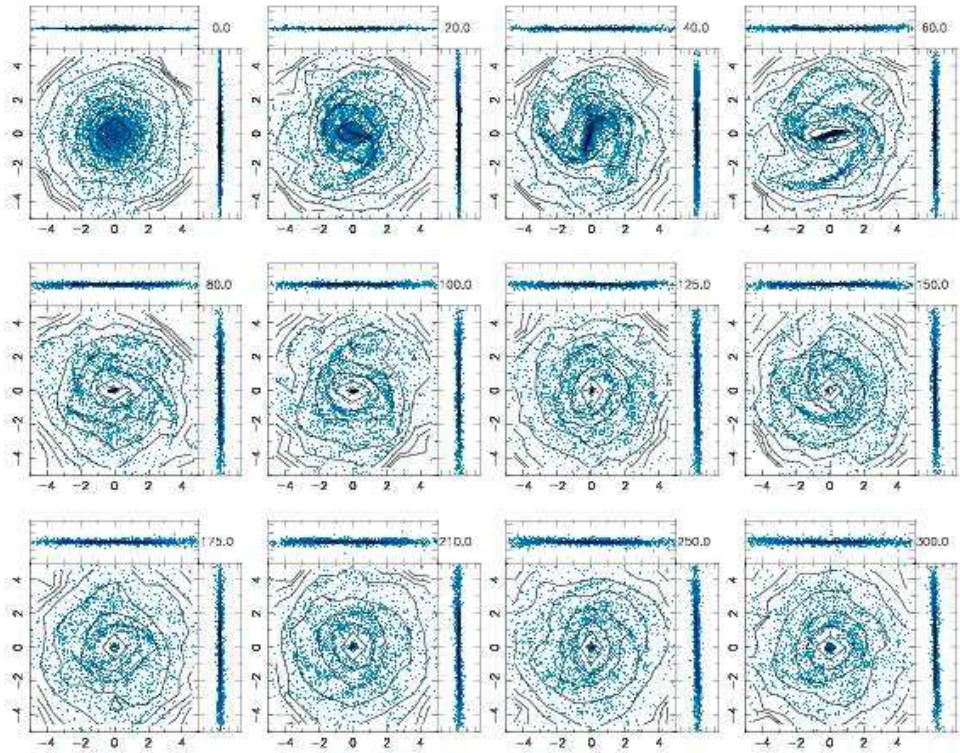}
 \caption{Evolution of the stellar and gaseous disc in the isolated
 barred galaxy model \Model{I}{0}. We show the face-on grey-scale
 density plot and the edge-on particle distribution of the gas. The
 stellar isodensity contours are shown face-on. The disc rotation is
 counter-clockwise. The time is given in the upper right corner
 of each panel in model units.}
 \label{fig02}
\end{figure*}

\begin{figure}
\begin{center}
 \includegraphics[scale=0.5,angle=-90]{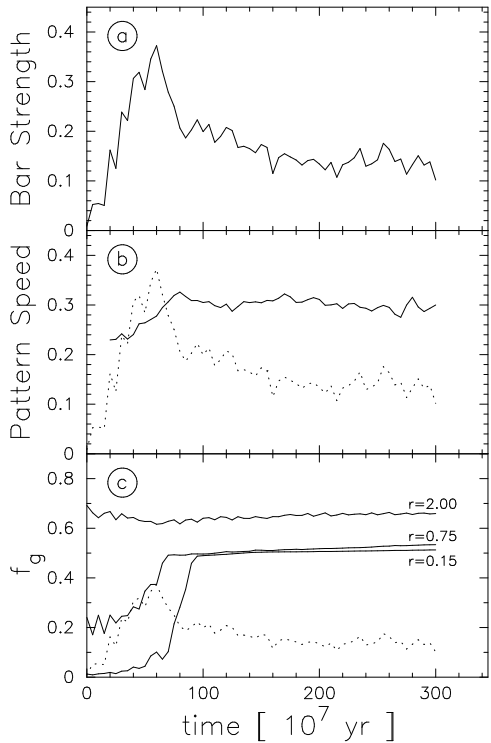} 
\end{center}
 \caption{The time evolution of a) the bar strength, b) the pattern
    speed and c) the gaseous mass fraction ($f_g$) within a constant
    radius in the isolated model \Model{I}{0} (full lines).  In
    panels b) and c) the bar strength is plotted with a dotted line
    for better comparison.}
 \label{fig03}
\end{figure}

 The morphological evolution of the stellar and gaseous disc is
 shown in Fig.~\ref{fig02}. The model of the disc galaxy was
 constructed so as to be globally unstable to non-axisymmetric
 perturbations and form a large-scale bar in a few disc rotations.
 At $t\!=\!60$, or after some $2\,t_{\rm rot}$, the stellar bar
 reaches its maximum strength, defined by the normalised amplitude
 of the $m\!=\!2$ Fourier component of the stellar disc particle
 distribution inside a cylindrical radius of 3.75\,kpc and one
 scale-height of the disc plane.  Care was taken in all models not
 to include any power from spiral features. The result is shown in
 Fig.~\ref{fig03}a.  At the time of its maximum strength the bar
 has a major axis length of $a\!=\!6$\,kpc and an axial ratio of
 approximately $3\!:\!1$, as determined by simple measurements of
 the stellar isodensity contours.
 Both stellar and gaseous trailing spiral arms emerge from the end
 of the bar. While the stellar arms slowly dissolve and are hardly
 visible by $t\!=\!100$, large spiral features in the gas persist
 throughout the run. The spiral pattern in the gas evolves from a
 two-arm to a flocculent-type spiral after the stellar arms have
 dissolved.  The gas also forms straight off-set shocks at the
 leading edge of the stellar bar. The shape and position of the
 shock loci indicate that the Lagrangian radius $r_{\rm L}$ is small
 and that the resonant $x_{\rm 2}$ orbit family is either absent or
 negligible at this early stage of evolution \cite{ath92b}.
 
 As soon as the bar forms, a substantial gas inflow sets in due to
 its gravitational torque, and within some $40\,\tau$, or
 approximately two bar rotations, some 50\% of the total gas, or
 $4.8\cdot 10^9$\,M$_\odot$, is driven towards the centre
 (Fig.~\ref{fig03}\,c) and accumulates there in a dense oval nuclear
 disc. This disc is elongated along the bar major axis and has a
 mean radius of approximately $r\!=\!0.36$\,kpc.  Although the
 nuclear disc appears to be aligned with the bar major axis, its
 shape and orientation could well be affected by the numerical
 softening, since its radius is a little less than half the softening
 length $\epsilon$. The mass of the nuclear disc represents some
 18\% of the total dynamical mass within a radius of 1.0\,kpc.  As a
 result of the growing mass concentration at the centre, the
 amplitude of the stellar bar decreases rapidly (Heller \&
 Shlosman 1994).  At $t\!=\!80$ the initial burst of inflow slows
 down and the stellar bar becomes quasi-stable (see
 Fig.~\ref{fig03}\,a), decaying at a much
 slower rate of $3\cdot10^{-4}\,\tau^{-1}$, as estimated by a linear
 least-squares fit. Likewise, owing to the inflow of gas to the
 centre, the pattern speed $\Omega_{\rm p}$ of the bar, as given by
 the phase angle of the $m\!=\!2$ Fourier component, increases
 linearly (Heller \& Shlosman 1994) till $t\!=\!80$
 (Fig.~\ref{fig03}\,b). After the central inflow of gas stops, i.e.
 at around $t\!\approx\!90$, the bar is rotating uniformly with a
 constant rate of $\Omega_{\rm p}\!=\!0.3\,\tau^{-1}$, or
 29 km\,s$^{-1}$\,kpc$^{-1}$, throughout the rest of the simulation.

\begin{figure}
 \begin{center}
   \includegraphics[scale=0.3,angle=-90]{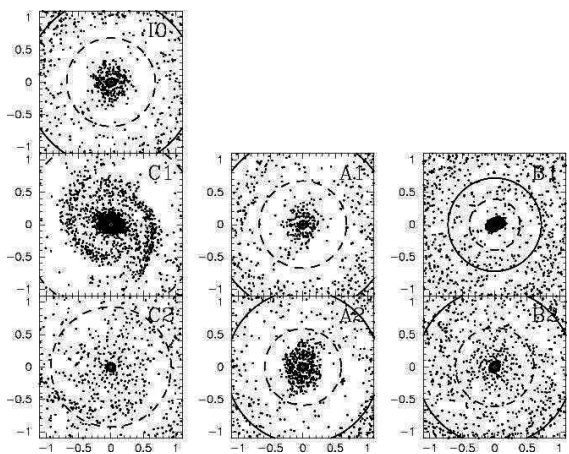}
 \end{center}
 \caption{Face-on plot of the gas particle distribution in the
     central disc region. The panels show the final gaseous mass
     distribution at the end of the different simulations (see
     Table~2), identified in the upper right corner of each panel.
     The circles mark the radius of the resonances obtained
     from the linear analysis: oILR (dashed lines) and UHR (full
     lines).}
\label{fig04}
\end{figure}

 The gaseous nuclear disc (hereafter n-disc) which has formed by
 the burst of gas inflow at first remains connected with the outer
 disc by two trailing spiral segments, which by way of their shocks
 feed material inward and thus contribute to its growth. Initially
 the nuclear disc is oval in shape, but as the bar weakens it
 becomes more circular. After $t\!=\!95$ the nuclear disc does not
 grow in mass anymore and the still inflowing gas starts to
 accumulate in a circumnuclear disc (hereafter cn-disc) surrounding
 the nuclear disc. This elongated cn-disc is orientated
 perpendicular to the bar major axis and has a lower surface density
 than the nuclear disc. In addition the cn-disc is more extended in
 size than the n-disc and has a radius of $r_{\rm cnd} = 0.9$\,kpc
 at the end of the simulation. The final morphology of the nuclear
 region is shown in Fig.~\ref{fig04}.
 The orientation of both the nuclear and the circumnuclear disc
 relative to the stellar bar indicate that the gas populates 
 the main families of periodic orbits in the barred potential (e.g.,
 Contopoulos \& Papayannopoulos 1980; Athanassoula et al. 1983).
 The $x_1$ orbits, which support the bar, are elongated along the
 bar major axis, while the $x_2$ orbits between the inner and outer
 Inner Lindblad Resonances (hereafter: iILR and oILR, respectively)
 are orientated perpendicular to the bar \cite{com}.

 In order to identify the presence and location of the main planar
 resonances in the disc at different times we average the mass
 distribution azimuthally and derive the
 $\Omega_{\rm c}\!-\!\kappa/2$ curve, where $\Omega_{\rm c}$ and
 $\kappa$ denote the circular and the epicyclic frequency,
 respectively. At the end of the run ($t\!=\!300$) this curve shows
 a pronounced maximum at a radius of $r\!=\!0.5$\,kpc exceeding the
 pattern speed $\Omega_{\rm p}$ of the bar, indicating the presence
 of ILRs. The resonant radii, obtained with this approximation, are
 given in Table~\ref{reso}.

\begin{table}
\caption{Resonances in the disc at different model times}
\label{reso}
\begin{tabular}{|r|c|c|c|c|c|c|c|c|} 
 time & model & $\Omega_p^{end}$ & i\,ILR &
 o\,ILR & UHR & CR & OLR \\
  \\[2.0ex] 
   60 & \Model{I}{0} & 0.28 & 0.11 & 0.23 & 1.40 & 2.16 & 3.85 \\
  150 & \Model{I}{0} & 0.30 & 0.03 & 0.64 & 1.25 & 1.99 & 3.48 \\[1.5ex]
  210 & \Model{I}{0} & 0.30 & 0.03 & 0.64 & 1.27 & 2.01 & 3.52 \\
      & \Model{C}{1} & 0.11 & 0.02 & 1.37 & 2.26 & 3.52 & 4.78 \\
      & \Model{A}{1} & 0.28 & 0.04 & 0.67 & 1.38 & 2.12 & 3.71 \\
      & \Model{B}{1} & 0.53 & 0.09 & 0.39 & 0.72 & 1.09 & 1.90 \\[1.5ex]
  300 & \Model{I}{0} & 0.29 & 0.03 & 0.68 & 1.31 & 2.08 & 3.68 \\
      & \Model{C}{2} & 0.16 & 0.01 & 0.93 & 2.22 & 3.63 & 4.93 \\
      & \Model{A}{2} & 0.28 & 0.02 & 0.59 & 1.18 & 1.93 & 3.47 \\
      & \Model{B}{2} & 0.33 & 0.02 & 0.61 & 1.20 & 1.85 & 3.14 \\ 
\end{tabular}
\end{table}

\begin{figure}
 \begin{center}
   \includegraphics[scale=0.5,angle=-90]{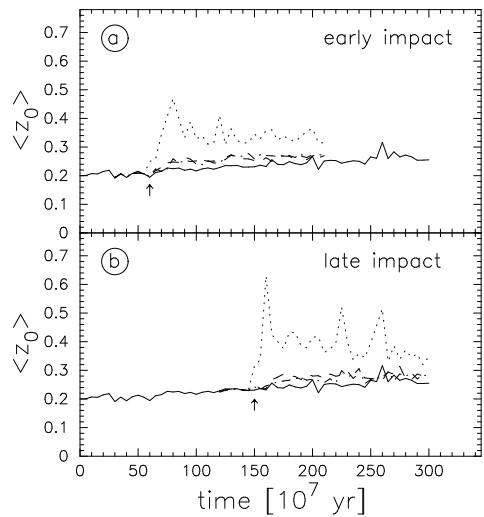}
 \end{center}
 \caption{Time evolution of the radially averaged scale-height $z_0$
    of the stellar disc.  The full line in both panels shows the
    isolated model \Model{I}{0}. In panel a) the evolution of the
    models \Model{C}{1}, \Model{A}{1} and \Model{B}{1} with an early
    impact is plotted with dotted, dashed, and dash-dotted lines,
    respectively. The same for panel b) for the late impact models
    \Model{C}{2}, \Model{A}{2} and \Model{B}{2}. The impact times
    are marked by an arrow in both panels.}
\label{fig05}
\end{figure}

 At its maximum strength, however, the stellar bar provides a strong
 non-linear perturbation to the gravitational potential of the disc
 and applying linear theory to identify the dynamical resonances is
 not sufficient.  A more reliable way is to search for the existence
 of the main orbit families \cite{ath92a}. For this we constructed
 surfaces of sections (SOS) by integrating orbits of a given
 Jacobian energy E$_{\rm J}$ in the equatorial plane, marking the
 points in the $(y,\dot{y})$ plane each time the orbits cross the
 line $x\!=\!0$ with $\dot{x}<0$ (e.g., Binney \& Tremaine 1987).
 The gravitational potential of each snapshot has been calculated on
 a non-equally spaced Cartesian grid with  a size of 6 by 6\,kpc.
  From that the potential and its derivatives are evaluated by using
  a piecewise polynomial function represented by a tensor product of
 one-dimensional B-splines. The times chosen for the SOS,
 i.e. at impact time and at the end of each run, are given in 
 Table \ref{init2}.
 We were thus able to confirm the presence of the ILRs in all of
 the cases, except for models \Model{I}{O} ($t\!=\!60$) and
 \Model{C}{1} ($t\!=\!210$).

\begin{figure*}
 \includegraphics[scale=0.68]{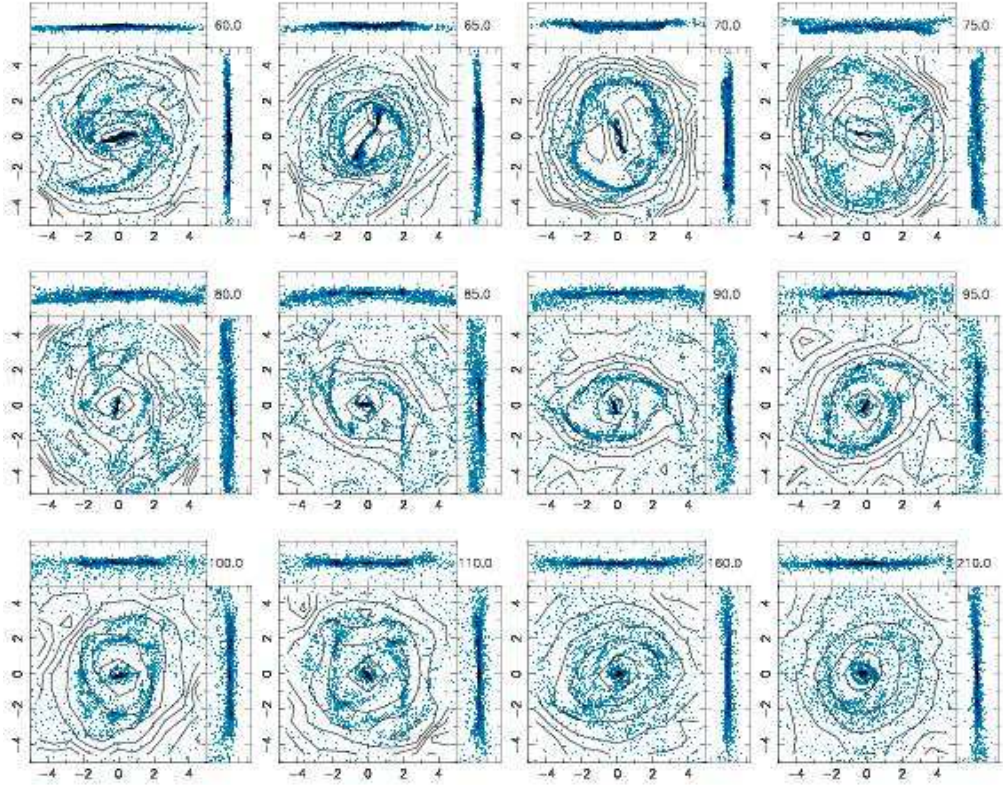}
 \caption{Evolution of the gaseous and stellar disc in the early
    central passage model \Model{C}{1}. The layout is as for 
    Fig.~\ref{fig02}.}
\label{fig06}
\end{figure*}
\begin{figure*}
\includegraphics[scale=0.68]{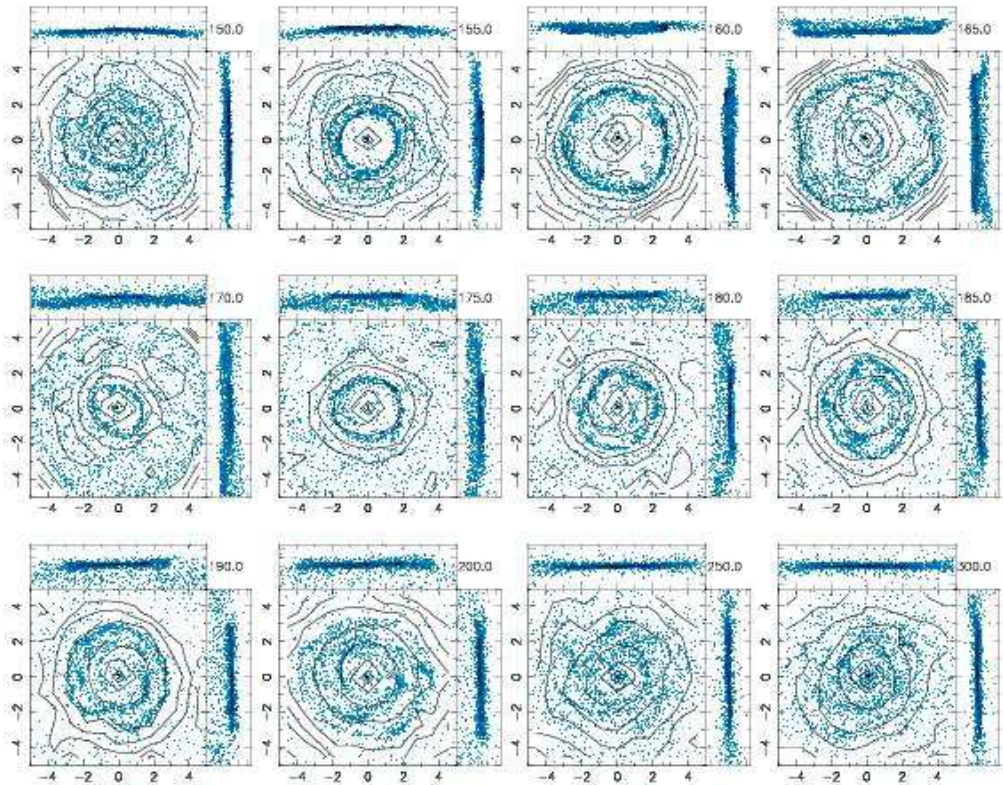}
 \caption{Same as Fig.~\ref{fig06}, but for the late central passage model
  \Model{C}{2}.}
\label{fig07}
\end{figure*}

 Most of the gas initially residing within corotation is driven
 towards the centre, resulting in a noticeable deficiency of gas
 outside the cn-disc, which could be interpreted as representing an
 HI hole. In contrast to this hole an oval ring of gas accumulates
 near the end of the weak stellar bar, close to the position of the
 ultra-harmonic resonance (UHR) at $r\!=\!2.25$\,kpc. After the
 resonance ring has formed ($t\!=\!175$) little further net radial
 redistribution of the gas  (less than 1\%) occurs.

 The stellar disc thickens vertically.  We follow the time evolution
 of its radially averaged vertical scale-height $z_0$ by fitting a
 sech$^2(z/z_0)$ density distribution to the particle distribution.
 The result is shown in
 Figure~\ref{fig05}. We find a linear increase of the vertical
 scale-height by a total of roughly 25\% in the isolated model.  We
 also notice a thickening of the gaseous disc, especially some
 slight flaring in the outer parts. Some of the thickening, however,
 might be induced numerically, either due to the limited number of
 particles in the model or to the relatively large softening. Since
 we use the same model later for the host galaxy, it will show the
 same numerical heating and any further change in scale-height may
 be attributed to the interaction.

 In this isolated model there is no perceptible sign of a
 peanut-shaped bar due to a buckling instability. This could
 be due to the fact that, even if the stellar disc were susceptible
 to such an instability, the presence of the gas would greatly
 dampen it \cite{bhsf}. 

\subsection{Interacting models} 

 The two times chosen for the impact of the companion correspond to
 the time (1) when the bar is at its maximum strength
 ($t_{\rm imp}\!=\!60$) and (2) when the bar is weak and has settled
 down to a quasi-stable state ($t_{\rm imp}\!=\!150$). Both times
 represent a characteristic epoch in the evolution of the disc. In
 the first case the bar has already formed and dynamically dominates
 the disc, but has still not had time to modify significantly the
 radial mass distribution of stars and gas. In the second case the
 stellar bar has weakened and the dynamical influence of the bar can
 be neglected compared to the effects of the interaction alone.  The
 impact position, i.e. the position of the centre of mass of the
 companion in the equatorial $z=0$ plane of the disc at $t_{\rm imp}$,
 also has been varied with respect to the bar.  In the models with
 major axis impacts, the impact position has a distance from the 
 centre of $r_{\rm imp}=6.0$\,kpc, which corresponds roughly to the
 corotation radius in the isolated model.  For the minor axis impacts,
 the distance is $r_{\rm imp}=3.0$\,kpc. 
 
\subsection{Central passage : Models \Model{C}{1} and \Model{C}{2}}

 In the simulations described in this section the companion hits the
 disc of the host galaxy at its centre. The morphological evolution
 of the models \Model{C}{1} ($t_{\rm imp}\!=\!60$) and \Model{C}{2}
 ($t_{\rm imp}\!=\!150$) is shown in Figures~\ref{fig06} and
 \ref{fig07}, respectively. Prior to impact both models show an
 axisymmetric vertical bending in both the stellar and the gaseous
 disc, with the inner regions pulled out in the direction of the
 approaching companion, as illustrated for the gaseous disc in the
 figures at $t\!=\!60$ and $t\!=\!150$ for models \Model{C}{1} and
 \Model{C}{2}, respectively.
 
\begin{figure}
 \begin{center}
   \includegraphics[scale=0.22,angle=-90]{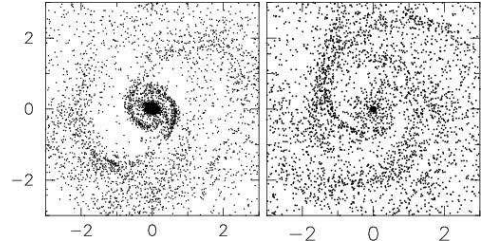}
 \end{center}
 \caption{Face-on plot of the final gas particle distribution in 
      the inner disc region of models \Model{C}{1} (left panel)
      and \Model{C}{2} (right panel) at the end of the simulation.}
\label{fig08}
\end{figure}

\begin{figure}
 \begin{center}
   \includegraphics[scale=0.25,angle=-90]{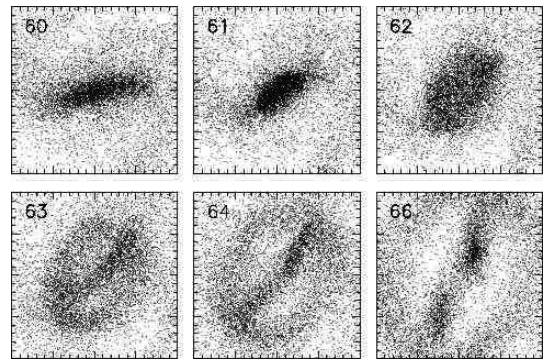}
 \end{center}
 \caption{Evolution of the stellar bar in model \Model{C}{1} after
 the impact. We show the face-on particle distribution of the
 stellar disc. The corresponding times in model units are given in
 the upper left corner of each frame and the size of the box
 corresponds to 12\,kpc.}
\label{fig09}
\end{figure}

\begin{figure}
 \begin{center}
   \includegraphics[scale=0.28,angle=-90]{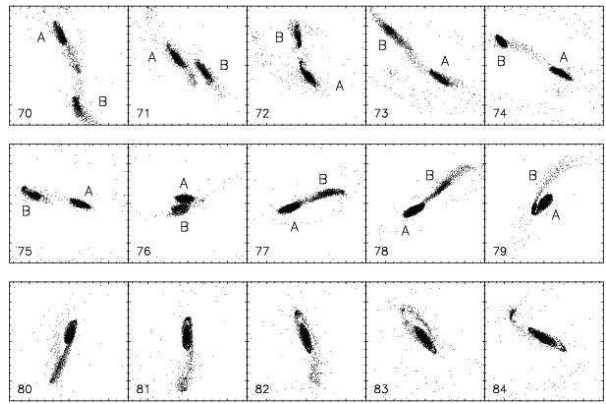}
 \end{center}
 \caption{Evolution of the gaseous disc in model \Model{C}{1} after
 the impact. We show the face-on gas particle distribution. The
 size of one box corresponds to 4.5\,kpc. The two
 gaseous fragments are labelled A and B and the time in model units
 is given in the lower left corner of each panel.}
\label{fig10}
\end{figure}
 
 The impact of the companion produces some expanding rings in the
 stars and gas, of which the latter are more sharply pronounced. The
 axial symmetry of the disc in model \Model{C}{1} is broken by the
 strong stellar bar and therefore the induced rings do not maintain
 their circular shape at larger radii (Fig.~\ref{fig06}, after
 $t\!=\!70$). Between the first and the second ring form several
 spokes, which are more pronounced in the gas, while their stellar
 counterparts are hardly visible by eye. All these induced features
 persist for only a few dynamical times after which the disc settles
 down once again to a quasi-steady state.  Although the stellar bar
 gets destroyed in both simulations, its dynamical imprint remains
 in the gas: namely the nuclear and circumnuclear discs, and the
 inner pseudo-ring. These components become visible again shortly
 following the impact and survive in the disc till the end of the
 runs (see Fig.~\ref{fig08}).

\subsubsection{Evolution of the bar}
 
 In the early impact model \Model{C}{1} both the stellar bar and the
 gas, which had accumulated prior to the impact in the shock loci
 inside the bar, get torn apart by the impact into two separate
 fragments. Fig.~\ref{fig09} shows this evolution for the stellar
 bar. When the companion passes through the disc, it exerts an extra
 inwards gravitational force on the disc particles and causes their
 orbits to contract. This affects the main orbit families of the
 bar and therefore the bar temporarily shortens somewhat
 (Fig.~\ref{fig09}, $t\!=\!61$).  After the companion has left
 the disc there is a strong rebound of the orbits, resulting in a
 radially expanding density wave.
 With the expansion of the particle orbits, the bar gets torn
 apart and the extremities of the two stellar fragments remain
 temporarily connected with the expanding stellar ring. At all
 times the two bar fragments are enclosed by the expanding ring.
  When the ring reaches a radius of approximately $r\!=\!6$\,kpc,
 the two bar fragments detach from it and sink back towards
 the centre, where they finally merge some $\Delta t\!=\!10$,
 or 0.1\,Gyr, after the impact and form a dense, almost axisymmetric
 centre.  Likewise, the gas concentrations at the
 shock loci inside the bar get torn apart, following the movement of
 the stellar bar fragments. As shown in Fig.~\ref{fig10} the two
 separated gas fragments flow back and forth inside the bar potential
 being trapped around $x_1$-like orbits. During this period they
 appear as two separate nuclei, before they finally merge and form
 a single nuclear disc some 0.2\,Gyr after the impact.

 In the later impact model \Model{C}{2} the stellar bar
 already has been weakened significantly due to the gas inflow.
 In this case also the impact produces expanding rings in the disc,
 but no stellar or gaseous fragments as in model \Model{C}{1}.

 In both models the stellar bar gets destroyed by the interaction
 almost immediately after the impact.  The bar strength as a
 function of time is shown in Fig.~\ref{fig19} (full line). In
 model \Model{C}{2} the already weak bar gets destroyed
 when the first stellar ring detaches from the central region.
 The temporary increase of the $m\!=\!2$ amplitude after the impact
 in model \Model{C}{1} results from the two stellar fragments which
 contribute to a bisymmetric
 distribution, but does not represent a stronger bar.
 No further stellar signatures of the former bar are left in the disc
 at the end of the runs.

\subsubsection{Rings and spokes}

 The passage of the companion through the disc excites both
 radial and vertical oscillations in the disc. The induced
 radial oscillatory motion of the particles produces 
 expanding density ring-waves (e.g., Lynds \& Toomre 1976)
 centred on the impact position, while the vertical oscillations
 lead to an increase of the vertical velocity dispersion and 
 a significant thickening of the disc.

 As shown in Fig.~\ref{fig06} and \ref{fig07}, an expanding
 ring is produced by the impact in both the stars and the gas. The
 ring first becomes visible in the gas, since the stellar ring is much
 broader and only becomes visible at sufficiently large radii.
 As the ring expands outward it becomes broader and its amplitude
 in both stars and gas decreases slowly as predicted by the impulse
 approximation (e.g., Binney \& Tremaine 1987).
 The ring in model \Model{C}{1} becomes asymmetric (Fig~\ref{fig06}, 
 $t\!=\!70$), since the underlying disc potential is perturbed by
 the presence of the stellar bar at least in the early phases of
 ring formation. In model \Model{C}{2}, however, the induced ring
 is more symmetric, since the perturbation due to the bar is weaker.
 In both models, the first ring reaches about 9\,kpc 
 at which point the inner part of the gaseous ring starts to
 fragment, with most of its mass flowing back towards the inner few
 kpc.

\begin{figure}
 \begin{center}
   \includegraphics[scale=0.37,angle=-90]{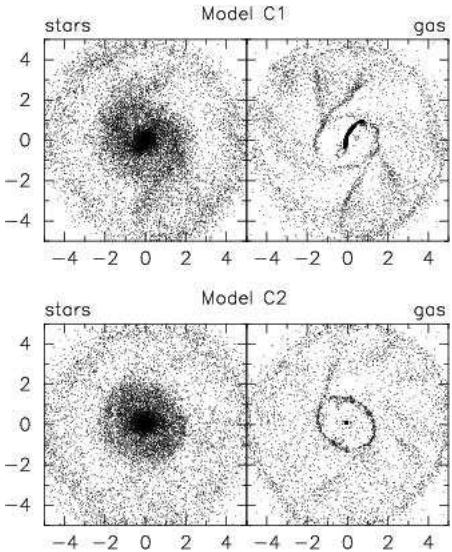}
 \end{center}
 \caption{Face-on particle distribution of models \Model{C}{1}
  and \Model{C}{2} at time  $\Delta t\!=\!20$ after the impact.
  The left and right frames show the stellar and gaseous disc,
  respectively. Only half the stellar particles are plotted for
  clarity. The spokes form between the first and second ring,
  the latter being visible mainly in the gas at this time. While
  the gaseous spokes are very prominent their stellar counterparts
  are less or hardly visible.}
 \label{fig11}
\end{figure}

 Following the first, a second expanding ring forms in both models,
 and becomes first visible in the gas after some $\Delta t\!=\!20$,
 or  $0.2$\,Gyr after the impact (see Fig.~\ref{fig11}).
 The second ring expands out to a radius of
 $r\!\approx\!6$\,kpc where it dissolves and its material
 is redistributed in the central region. In the stellar disc the second
 ring is hardly visible, since its amplitude is too low and it does
 not expand as far out as the first one.
 While only two rings form in the stars -- a third one can
 (hardly) be identified only in the radial density profile of the disc --
 we find several consecutive expanding ring features in the gas
 which are more asymmetric than the first two rings
 (e.g., Fig.~\ref{fig07} $t\!=\!190, 200$ and Fig.~\ref{fig14}
 $t\!=\!57$).

 The gas fragments that detach from the inner side of the first ring
 flow back to the central disc region and are sheared out by the
 differential rotation of the disc, forming several spokes
 between the two collisional rings (Fig.~\ref{fig11}).
 As noted by APB97, spokes form only between the first and the second
 ring, the presence of the latter being necessary.
 The properties of spokes in collisional ring galaxies are discussed
 in detail by Hernquist \& Weil \shortcite{hw93}.
 
\begin{figure}
 \begin{center}
   \includegraphics[scale=0.5,angle=-90]{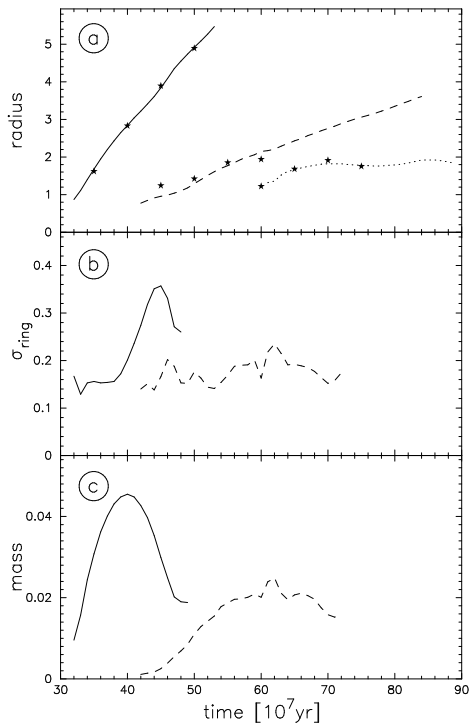}
 \end{center}
 \caption{Properties of the stellar and gaseous rings in model
  \Model{C}{2}. Panel a) shows the mean radius of the three gaseous
  (full, dashed and dotted lines, respectively) and stellar rings
  (marked by stars) as a function of time. Panel b) shows the width
  $\sigma$ of the gaseous rings obtained from a Gaussian fit. Panel
  c) shows the mass of these rings.}
\label{fig12}
\end{figure}

\begin{figure}
 \begin{center}
   \includegraphics[scale=0.4,angle=0]{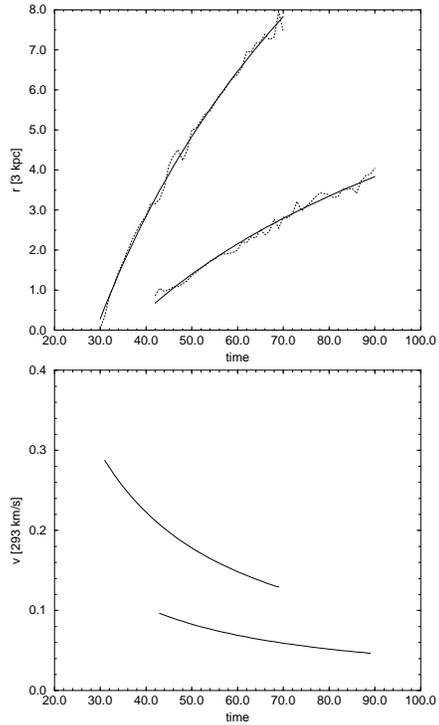}
 \end{center}
 \caption{Properties of the gaseous rings in model \Model{C}{2}.
   The upper panel shows the radius of the gaseous rings as a
   function of time. The `raw' data are given by dotted lines and
   the fitted function by full lines. The lower panel shows the
   derivative of the fitted function.}
\label{fig13}
\end{figure}

\begin{figure}
 \begin{center}
   \includegraphics[scale=0.28,angle=-90]{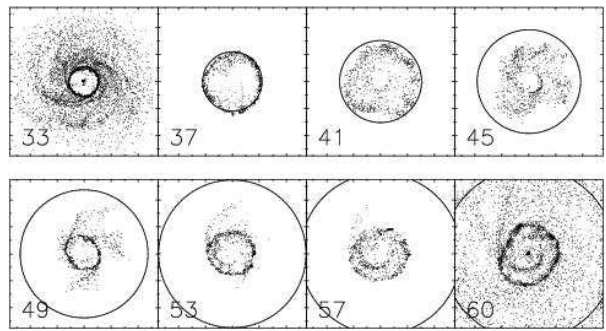}
 \end{center}
 \caption{Expansion of the gaseous rings in model \Model{C}{2}.
  The circle in the panels
 shows the position of the first ring. The time in model units is
 given in the lower left corner of each panel. The first and the
 last panel show the SPH particle distribution of the whole disc, 
 while the remaining panels include only the particles contributing
 to the first gaseous ring at $t\!=\!33$.}
\label{fig14}
\end{figure}

 To track the position of the rings and derive their expansion
 velocity, we determine the azimuthally averaged density of annuli
 of varying radii in the disc. Then we fit a polynomial function
 to subtract the background of the disc and a Gaussian to fit the density
 distribution inside the ring. Using this method we get information
 about the position, width and mass of the rings. The results are
 shown in Figure~\ref{fig12}.
 We find that the gaseous ring traces the position of the stellar
 ring, i.e. its position coincides with that
 of the stellar ring (Fig.~\ref{fig12}a) and both expand with
 the same velocity.  With increasing radius the stellar and gaseous
 rings become broader (Fig.~\ref{fig12}b) and less dense, as
 predicted by the impulse approximation (e.g., Binney \& Tremaine 1987).

 By integrating the Gaussian with the fitted parameters we can
 estimate the gas mass contributing to the rings.
  Fig.~\ref{fig12}c shows the mass of the gaseous rings as a
 function of time. Unlike stars, the orbits of the gas cannot cross
 and up to $t\!=\!40$ gas is piled up in the ring as it expands
 outward. Due to shock dissipation and self-gravity of the ring,
 its inner parts start to fragment and thus its mass decreases.
 We fit a logarithmic function to the radii of the gaseous rings
 from which we derive their expansion velocity.  Fig.~\ref{fig13}
 shows the position and expansion velocity of the first and second
 ring in model \Model{C}{2}. As has been previously found
 (e.g. APB97) the expansion velocity of the rings decreases with
 radius, as predicted by the impulse approximation.
 The expansion velocity of the first ring drops gradually
 from 88 km\,s$^{-1}$ (7.5\,$v_{\rm s}$) to 35 km\,s$^{-1}$
 (3\,$v_{\rm s}$, where $v_{\rm s}$ is the sound speed in the gas).
 The second ring starts with a lower velocity, which decreases from
 29 km\,s$^{-1}$ (2.5\,$v_{\rm s}$) to 11.72 km\,s$^{-1}$ (1\,$v_{\rm s}$).
 
\begin{figure}
 \begin{center}
   \includegraphics[scale=0.3]{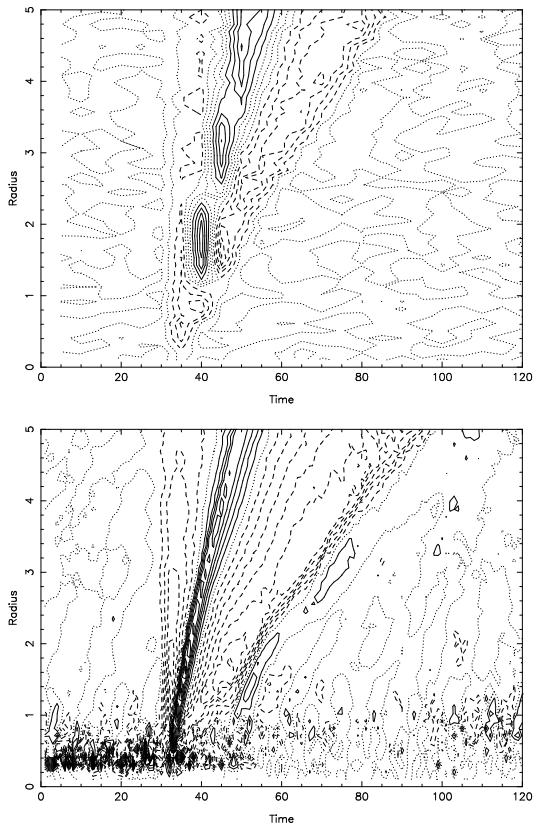}
 \end{center}
 \caption{Isocontour plots of the radial velocity for the stellar
   (upper panel) and gaseous (lower panel) disc. Dotted contours
   correspond to -10, 0 and 10 km\,s$^{-1}$, solid line contours
   to 20, 30, 40, ...  km\,s$^{-1}$ and dashed ones to -20, -30,
   -40, ... km\,s$^{-1}$.}
\label{fig15}
\end{figure}

 In Fig.~\ref{fig15} a and b we show the isocontours of the radial
 velocities for the stellar and gaseous disc, respectively, as a
 function of both radius and time. It shows the radial osciallation
 of the disc material, first inward following the impact, followed
 by an outward movement as the ring passes through, and finally again
 inward.  The amplitude of this motion can be seen to be
 considerable.  With each passage of a ring the local radial
 velocity dispersion (not shown here) in the disc increases, until
 further density enhancements of this kind are no longer supported.

 The passage of the companion also gives rise to vertical
 oscillations in the disc (e.g., Weinberg 1991;
 Mihos \& Hernquist 1994) giving it a layered appearance in the
 edge-on projection (see Fig.~\ref{fig06}, e.g., $t\!=\!75$ and
 Fig~\ref{fig07}, e.g., $t\!=\!165$).
 The same type of vertical motion can be seen in the numerical 
 models by Lynds \& Toomre \shortcite{lyt76} and also in
 Mihos \& Hernquist \shortcite{mih94}.
 In fact the layered appearance is a projection effect resulting
 from the different vertical position of the outer, still
 unperturbed, gas disc and the expanding ring.
 As a consequence of the vertical bending in the disc, induced by 
 the companion prior to impact, the first expanding ring starts
 forming above the equatorial $z\!=\!0$ plane (Fig.~\ref{fig07},
 $t\!=\!155$). At larger radii, i.e. at later times, the ring
 crosses the $z\!=\!0$ plane of the disc (defined by the centre of
 mass of the gaseous disc) driven by the passage of the
 companion.  With each passage of a ring through the equatorial
 disc plane the vertical velocity disperion increases and results in
 a significant thickening of the disc. The vertical scale-height 
 of the stellar disc is shown in Fig.~\ref{fig05}. We find an abrupt
 increase of $<\!z\!>$ from a value of roughly 0.6\,kpc to 1\,kpc at 
 $\Delta t\!\sim\!15$ after the impact.

\begin{figure*}
 \includegraphics[scale=0.65]{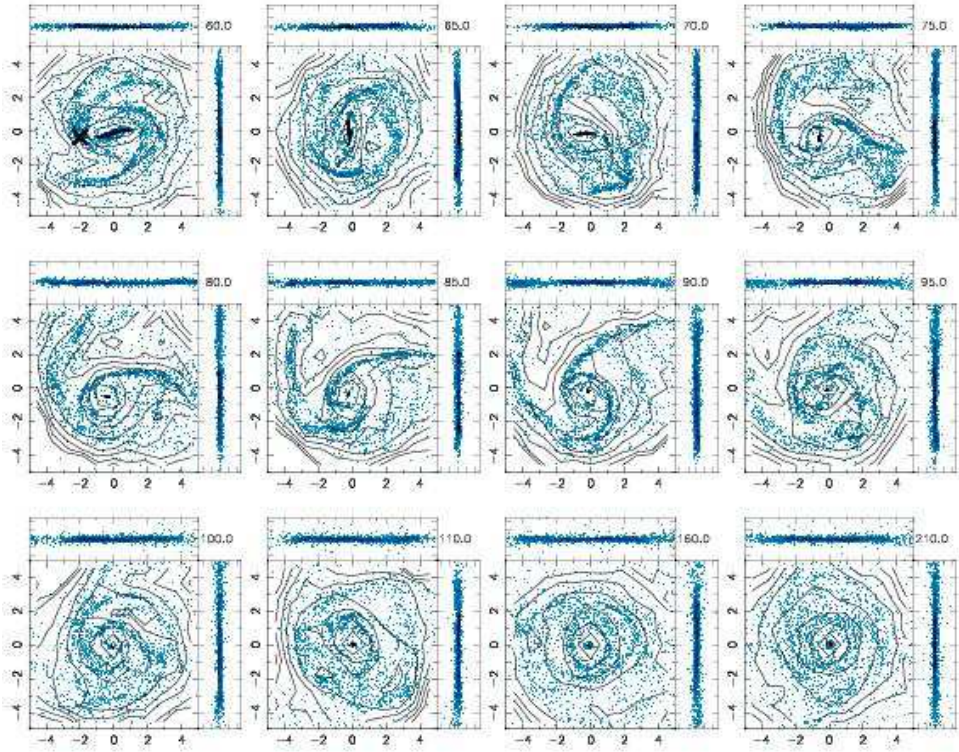}
 \caption{Evolution of the stellar and gaseous disc in model
   \Model{A}{1}, where the impact occurs on the bar major axis at an
   early time. The layout is as for Fig.~\ref{fig02}.
   The origin in the frames is always centred on the centre of
   mass of the halo component and the impact position of the companion
   is marked by a cross in the first frame.}
 \label{fig16}
\end{figure*}
\begin{figure*}
\includegraphics[scale=0.68]{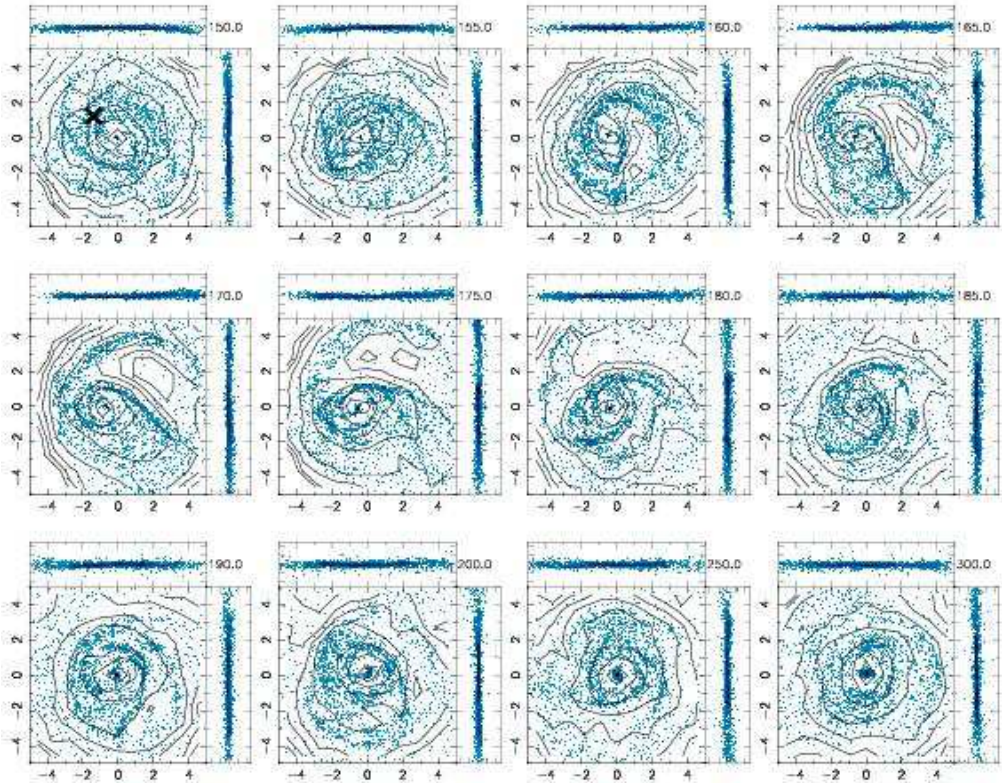}
 \caption{Same as Fig.~\ref{fig16} but for the late impact model
          \Model{A}{2}.}
 \label{fig17}
\end{figure*}

 To analyse the radial and vertical oscillations in the disc in 
 more detail, we trace the orbits of individual particles in $r$
 and $z$ as a function of time. With a Fourier transformation
 we find that the principal frequencies correspond to the
 radial and vertical epicyclic frequencies in the disc. The
 increase of kinetic energy in $z$ is therefore responsible
 for the heating or thickening of the disc after the impact.

\subsubsection{Nuclear and circumnuclear disc}

 In  model \Model{C}{1} (strong bar) the two gas nuclei, which 
 have formed after the interaction in the disrupted bar, merge
 and form a single nuclear disc. This nuclear disc, however, is
 less massive and more extended than the one in the isolated model.
 After the n-disc has formed, it remains connected by two gaseous
 trailing spiral segments with the second induced ring ($t\!=\!85$
 in Fig.~\ref{fig06}).
 The gas which is driven towards the centre continuously by way of
 the spiral shocks accumulates in a circumnuclear disc.
 Compared to the isolated model, we find about 5\% more gas
 in the cn-disc inside $r\!=\!2.0$\,kpc.  We also find some two-armed
 trailing spiral structure of varying strength in the cn-disc.
 At the end of the run of model \Model{C}{1}, smoothly distributed
 gas remains in the outer disc, as well as in a dense nuclear
 disc/core (like in the isolated model) surrounded by a circumnuclear
 disc with a radius of $r\!=\!3$\,kpc (Fig.~\ref{fig04}). This disc
 is orientated almost perpendicular to the nuclear disc and is connected
 by two trailing spiral arms with an outer distorted ring which has
 formed at a radius of $r\!\approx\!6$\,kpc, close to the UHR of the
 former bar (see Fig.~\ref{fig08}).

 In model \Model{C}{2} the gaseous nuclear disc has already formed
 before the impact. With the companion passing through the disc, the
 inner disc region is pulled out of the $z\!=\!0$ disc plane towards
 the direction of the companion.
 The gaseous nuclear disc reaches a maximum vertical distance relative
 to the (first) expanding ring of approximately $\Delta z\!=\!1.5$\,kpc
 at $\Delta t\!=\!8$, or $8\cdot10^7$\,yr, after the impact. After the
 collisional rings have dissolved and the n-disc has recentered
 vertically, more gas is driven towards the centre by way of the
 spiral shocks which have newly formed in the surrounding disc.
 The radius of the nuclear disc is larger than in the isolated model
 at the same time, but has about the same mass. 
 The inflowing gas accumulates in a circumnuclear disc, which has
 a mean radius of $r\!\approx\!2.1$\,kpc at the end of the simulation
 and is connected by two spiral arms with the outer ring,
 which has formed close to the UHR. The region between
 the cn-disc and the ring at the UHR (at roughly $r\!\approx\!6$\,kpc) shows
 a clear deficiency of gas, which also exists in model \Model{C}{1}.

 After the bar has been destroyed, the gaseous nuclear discs in both
 models maintain their slightly elongated shape and still rotate
 with the same pattern speed as the former stellar bar before
 the impact.
 In both models we find spiral shocks in the circumnuclear disc
 (Fig.~\ref{fig04}), which persist till the end of the
 simulation.

\subsection{Major axis passages : Models \Model{A}{1} and
    \Model{A}{2} }

 In these simulations the companion hits the disc on the bar major
 axis at a radius of approximately 6.0\,kpc, i.e. in model
 \Model{A}{1} close to the end of the strong bar and close to the
 corotation radius as determined from the isolated model. The
 morphological evolution of these models is shown in
 Fig.~\ref{fig16} and \ref{fig17} for the early (\Model{A}{1}) and
 late (\Model{A}{2}) impact, respectively.

 When the companion approaches the disc it exerts an extra
 gravitational force on the disc particles and the bar gets shifted
 towards the impact position, almost merging with the spiral arm
 close to the impact position (Fig.~\ref{fig16}, $t\!=\!65$). While
 the central impacts produce a closed expanding ring structure, the
 off-centered passages produce long spiral arms, which do not form a
 closed ring (see Fig.~\ref{fig16}, $t\!=\!70$ and Fig.~\ref{fig17},
 $t\!=\!165$), in good agreement with the results of
 Toomre \shortcite{too78} for non-barred galaxies.
 
\subsubsection{Evolution of the bar}
  
\begin{figure}
 \begin{center}
   \includegraphics[scale=0.5,angle=-90]{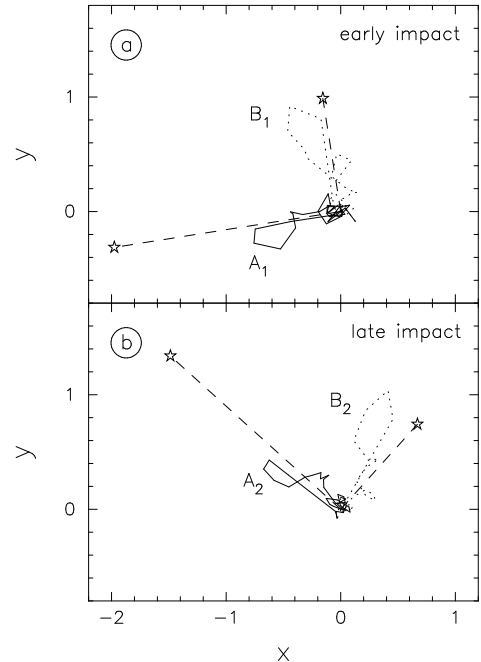}
 \end{center}
 \caption{Bar displacement. Panels a) and b) show the relative
 distance between the centre of mass of the bar and that of the
 halo for the early and late impact models, respectively. The
 stars and dashed lines in each panel mark the impact location
 in the disc.  The models for major and minor axis impacts are
 shown with full and dotted lines, respectively.}
\label{fig18}
\end{figure}

\begin{figure}
 \begin{center}
   \includegraphics[scale=0.5,angle=-90]{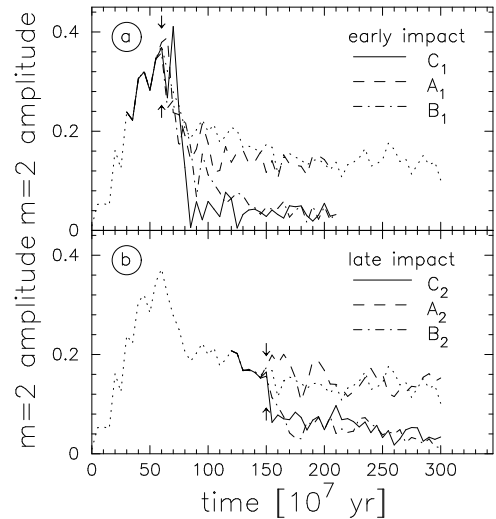}
 \end{center}
 \caption{Bar strength as a function of time. Panel a) and b) show
  the $m\!=\!2$ component for the models with early and late
  impacts, respectively. The isolated model is shown in both panels
  (dotted line).  The impact time is marked with arrows.}
\label{fig19}
\end{figure}

 Immediately following the passage of the companion through the
 disc plane, the stellar bar gets displaced from the centre to
 a maximum radius of 2.4\,kpc in the direction of the impact
 position in both models \Model{A}{1} (early impact) and \Model{A}{2}
 (late impact). The off-centering of the bar in these types of
 interaction has already been described by several authors (e.g., 
 Gerber \& Lamb 1994, 1996; Athanassoula 1996a; APB97).
 To measure the displacement of the bar we trace its centre of mass
 iteratively within a cylindrical shell of constant radius and
 then determine the distance to the centre of mass of the halo,
 which we can assume to be the dynamical centre of the galaxy.
 Figure~\ref{fig18} shows the relative distance of the centre of
 mass of the bar from this centre in the galactic plane as a
 function of time. In both models, the bar, still rotating around
 its own centre, reaches its maximum distance at $\Delta t\!=\!10$
 or 0.1\,Gyr after the impact and recenters over a period of some
 2 disc rotations, or 0.6\,Gyr. The shape of the stellar bar does
 not change significantly while being off-centered. The bar 
 strength is not affected by the interaction compared to the
 isolated model (Fig.~\ref{fig19}), although the bar also
 weakens due to gas inflow.

\begin{figure}
 \begin{center}
   \includegraphics[scale=0.5,angle=-90]{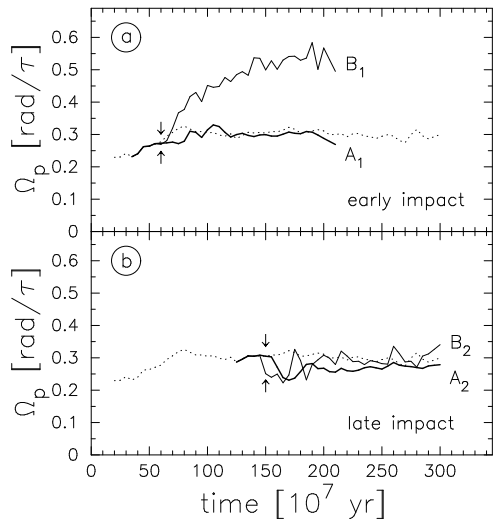}
 \end{center}
 \caption{Pattern speed $\Omega_p$ as a function of time.
  Panel a) shows model \Model{A}{1} (thick line) and \Model{B}{1}
  (thin line) with the early impact at $t\!=\!60$.
  The dotted
  line shows the evolution for the isolated model \Model{I}{0}. 
  The same for panel b) for models \Model{A}{2} (thick line) and
  \Model{B}{2} (thin line) with impact at time $t\!=\!150$.
  The impact times are marked by arrows in both panels.}
\label{fig20}
\end{figure}

\begin{figure}
 \begin{center}
   \includegraphics[scale=0.3,angle=-90]{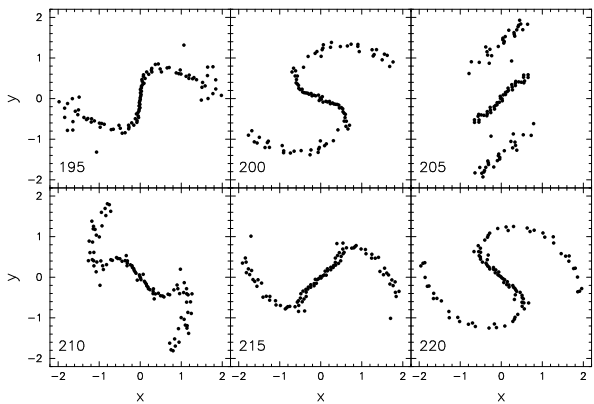}
 \end{center}
 \caption{Phase angle of the $m\!=\!2$ component of the stellar mass
  distribution as a function of radius for model \Model{A}{2}.
  Note that the outer spiral arms rotate slower than the central
  stellar bar. At $\Delta t\!=\!60$ ($t\!=\!210$) a leading spiral
  segment is also present, connecting the bar to the outer spirals.}
\label{fig21}
\end{figure}

 In model \Model{A}{1} the off-centered bar detaches from the ring
 due to its rotation and moves to the centre again. We find that
 the outer spiral arms in model \Model{A}{2} rotate temporarily
 with about half of the bar pattern speed for some 0.6 Gyr after
 the impact.  Fig.~\ref{fig21} shows the phase angle of the
 $m\!=\!2$ component of the stellar disc mass distribution as a
 function of radius for different times.  The stellar bar detaches
 from the spiral arms at about $\Delta t\!=\!55$ after the impact
 and reconnects again at $\Delta t\!=\!60$.  A similar effect has
 been described by Sellwood \& Sparke (1988) (see their Fig.~2)
 who studied the dynamics of an isolated galaxy with different
 pattern speeds for the bar and the spiral arms. In their models
 the bar detaches periodically from the spiral arms. In our
 simulation, however, the bar detaches only once from the spiral
 arms, because the spiral pattern speed increases again to its
 initial value of $\Omega\!=\!0.3\,\tau^{-1}$ after roughly
 $\Delta t\!=\!60$ after the impact.
 The pattern speed $\Omega_p$ of the bar in model \Model{A}{1}
 decreases after the interaction for some 0.3\,Gyr, but reaches its
 initial value of 0.3\,$\tau^{-1}$ when the bar has recentered (see
 Fig.~\ref{fig20}).  The spiral arms in this model rotate with the
 same pattern speed as the bar and remain connected to it at all
 times.

\subsubsection{Nuclear and circumnuclear discs}

 In model \Model{A}{1} the gas, which was before the impact
 in the shock loci inside the stellar bar, is driven
 towards the centre of the displaced bar and accumulates in a small
 elongated nuclear disc, which is aligned with the bar major axis.
 At $t\!=\!90$ an elongated gaseous ring starts to form with a
 constant mean radius of approximately $r\!=\!4.5$\,kpc, located
 close to the bar UHR. The ring and the nuclear disc are connected
 by spiral arms, which form at about $t\!=\!100$.  By way of the
 spiral shocks more gas is driven towards the centre and accumulates
 in a circumnuclear disc, which is slightly elongated perpendicular
 to the stellar bar.


 The evolution in model \Model{A}{2} is similar. The gaseous nuclear
 disc is already present before the impact and after the stellar bar
 has recentered, spiral shocks between the nuclear disc and a
 gaseous ring, which has formed close to the UHR, drive more gas to
 the centre.  This gas accumulates in a circumnuclear disc.
 In both models, \Model{A}{1} and \Model{A}{2}, we find a small gap
 between the nuclear and
 circumnuclear disc with a deficiency of gas. The final gas
 morphology of the central disc region in both models is shown in
 Fig.~\ref{fig04}.

\subsubsection{Rings and spiral arms}

\begin{figure}
 \begin{center}
   \includegraphics[scale=0.26,angle=-90]{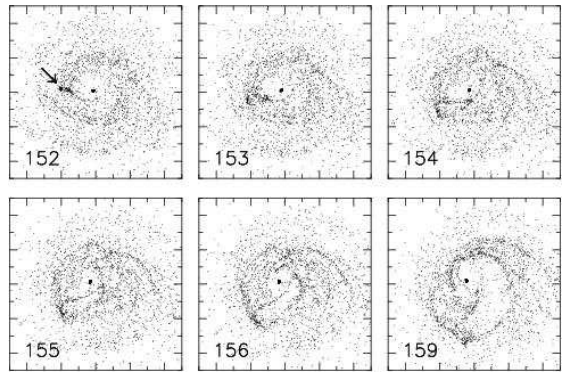}
 \end{center}
 \caption{Evolution of the gaseous disc in model \Model{A}{2} after
 the impact. We show the face-on gas particle distribution. The time
 in model units is given in the lower left corner in each frame. The
 size of each box corresponds to 30\,kpc. The position of the 
 forming density wave is marked in the first frame by an arrow.}
\label{fig22}
\end{figure}

 As in the central impact models, the passage of the companion
 through the disc produces a radially expanding density wave
 in both the stars and the gas, originating from the impact
 position. However, since the impact in model \Model{A}{1} and
 \Model{A}{2} is off-centered, the symmetry of the ring wave
 is broken as soon as it encounters the stellar bar and spiral
 arms in the disc. This is illustrated for model \Model{A}{2}
 in Figure~\ref{fig22}. An expanding ring-like structure forms
 at the impact position, but gets sheared out by the differential
 rotation in the disc as it expands, forming a {\em pocket}
 structure as mentioned by  Gerber \& Lamb (1994) who studied the
 caustics produced by an off-centered impact in a non-barred disc
 model. 
 Because the mass distribution in the disc is more symmetric than
 in model \Model{A}{1}, the features produced are more symmetric.
 Inside the ring or {\em pocket} there is a deficiency of gas.
 The density wave expands out more freely in the direction of
 disc rotation. When it encounters the spiral arm, it joins it to
 form one long tidal arm, which is connected to the end of the
 off-centered bar (see e.g. Fig.~\ref{fig16},
 $\Delta t\!\approx\!15$ after the impact).

 In both models a second expanding density wave starts to form at
 $\Delta t\!\approx\!20$ after the impact, starting from the
 impact position in the rotating frame of the disc.
 When the density wave merges with the spiral
 arm, a tidal arm is formed which is more linear than the first
 tidal arm that previously formed on the opposite side of the bar.
 By this time the first tidal arm in model \Model{A}{1} already
 extends out to a radius of $r\!=\!18$\,kpc, exceeding the initial
 disc radius.  Between the two tidal arms there is a clear
 deficiency of stars and gas. Both tidal arms wind up and dissolve
 slowly with most of their material redistributed in the inner
 15\,kpc again. Some material, however, is left in the outer disc,
 giving it an asymmetric shape.

\begin{figure*}
\includegraphics[scale=0.65]{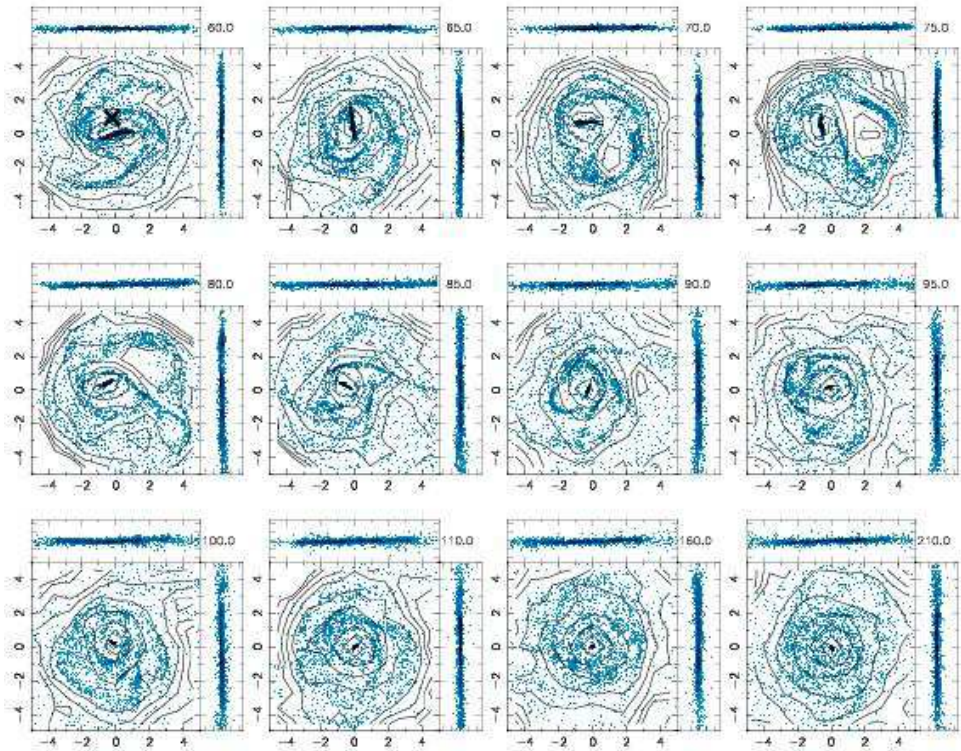}
 \caption{Evolution of the stellar and gaseous disc in model
   \Model{B}{1}, where the impact occurs on the bar minor axis at
   an early time. The layout is as in Fig.~\ref{fig16}.}
\label{fig23}
\end{figure*}

\begin{figure*}
\includegraphics[scale=0.65]{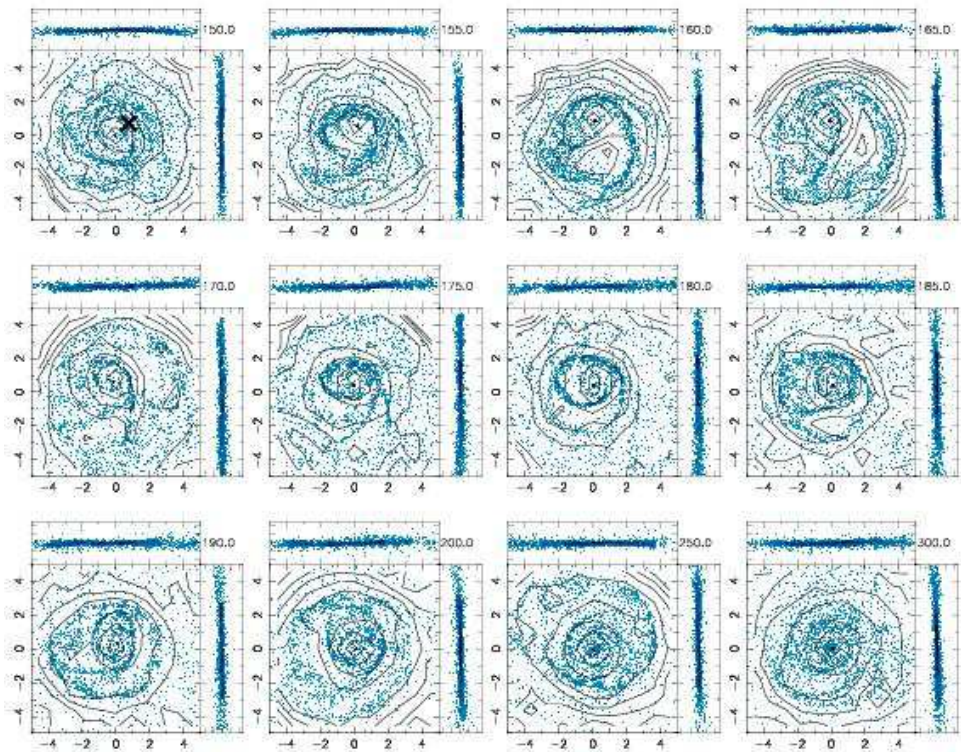}
 \caption{Same as Fig.~\ref{fig23} but for the late impact model
    \Model{B}{2}.}
\label{fig24}
\end{figure*}

\subsection{Minor axis passages : Models \Model{B}{1} and
    \Model{B}{2} }

 In this section we describe the evolution of the models in which
 the companion hits the disc in the direction of the bar minor axis
 at a distance of approximately $r_{\rm imp}\!=\!3.0$\,kpc, just 
 outside the bar. The impact times for models \Model{B}{1} and
 \Model{B}{2} are $t_{\rm imp}\!=\!60$ and $t_{\rm imp}\!=\!150$,
 respectively. Figures \ref{fig23} and \ref{fig24} show the
 evolution of both the stellar and the gaseous disc.
 
 Before the impact the evolution of the disc is similar to the
 isolated model \Model{I}{0}. In contrast to the central impact
 cases we do not find a vertical bending of the disc prior to impact
 in either of the two models. After the impact the bar is
 temporarily off-centered, and the passage of the companion excites
 radial oscillations in the disc, resulting in the formation
 of an expanding density wave.  Due to its off-centering the impact
 produces two long tidal arms rather than a closed ring structure.
 Finally the stellar bar gets almost destroyed by the interactions
 in both models. 

\subsubsection{Evolution of the bar}

 Immediately following the impact, the stellar bar gets displaced
 from the centre in both models to a maximum distance of
 approximately 3.0\,kpc towards the impact position, i.e. to a
 distance comparable to the one of the impact location from the
 centre of the target disc (Fig.~\ref{fig18}). The off-centered
 bar still rotates around its own centre, while moving radially in
 the disc.  The bar reaches its maximum
 distance from the dynamical centre at a time of $\Delta t\!=\! 10$
 after the impact of the perturber, which is the same time-scale as
 in the major axis impact models. The offset of the bar lasts for a
 period of roughly $\Delta t\!=\!0.6$\,Gyr, or about two disc
 rotations, in both models, as for models \Model{A}{1} and
 \Model{A}{2}.
 
 We measure the strength and pattern speed of the bar with the
 method decribed in section~3.1. Care was taken that no power from
 spiral features was included in the computation
 by checking that the phase angle is constant with radius.
 As the bar strength weakened the error in the pattern speed
 measurement increased, but was always below about 5\%.
 While the pattern speed in model \Model{B}{2} remains constant
 after the impact, we find an increase of $\Omega_{\rm{p}}$ in the
 early impact model \Model{B}{1} from 0.3\,$\tau^{-1}$ to a rate of
 0.5\,$\tau^{-1}$ (see Fig.~\ref{fig20}), presumably due to the
 torque by the companion on the strong bar.  Since the rotation
 curve of the disc does not change significantly after the impact,
 the change of $\Omega_p$ is accompanied by a change of the
 positions of the resonances in the disc. In fact this change can be
 observed in the resonance gas ring located close to the bar UHR.
 While $\Omega_p$ increases, the UHR moves inward and the ring
 shrinks from a radius of approximately $r\!=\!3$\,kpc to 2.4\,kpc,
 compatible with the change of $\Omega_p$.

 By the end of the simulation the stellar bar is almost destroyed
 by the interactions in both models.  However, the dissolution
 time-scale for the early impact case \Model{B}{1} is longer than in
 the corresponding model with central impact (Fig.\ref{fig19}).
 There is an indication that this might also be true, but to a
 lesser degree, for the late impact case \Model{B}{2}. The 
 vertical scale-height of the disc, however, does not increase
 considerably after the interaction (see Fig.~\ref{fig05}).

\subsubsection{Nuclear and circumnuclear disc}
 
 In model \Model{B}{1}, where the bar is strong at impact time, the
 leading gaseous shocks remain inside the off-centered stellar bar.
 As the bar is recentering, the gas flows toward the centre of the
 bar. Following the bar recentering this gas forms a dense nuclear
 disc which is aligned with the bar major axis. Over time the
 nuclear disc looses its elongated shape and becomes more circular,
 while the inflow rate inside a central region of $450$\,pc remains
 constant at about $4\,\mbox{M}_{\sun}$\,yr$^{-1}$.
 The total amount of gas driven towards the centre, which is limited
 by the amount of gas within the bar region before the impact, is the
 same as in the isolated model. In this model, however, no
 circumnuclear gaseous disc forms (see Fig.~\ref{fig04}).

 The nuclear disc in model \Model{B}{2} has already formed before
 the impact and remains in the centre of the temporarily off-centered
 bar. Some of the inflowing gas accumulates in a large diffuse
 circumnuclear disc, almost extending out to the gas ring
 at 3\,kpc, which has formed following the impact.

\subsubsection{Rings and spiral arms}

 The passage of the companion through the disc produces an expanding
 density wave, that becomes visible first in the gas. The general
 evolution of this ring-wave is similar to the one in the major axis
 impact models. As can be seen in Fig.~\ref{fig23} and
 \ref{fig24} the density wave expands more freely in the direction
 of disc rotation and maintains longer its circular shape on that
 side ($\Delta t\!=\!5$ after the impact) before merging with the
 spiral arm in the outer disc. As soon as the wave encounters the
 spiral arm, the ring feature opens up and forms a long tidal arm,
 extending out to a maximum radius of approximately $r\!=\!18$\,kpc.
 Then the tidal spiral arm dissolves slowly and most of its material
 is redistributed in the initial disc region. The other part of the
 ring remains close to the off-centered bar and is connected to it
 by a straight spoke ($\Delta t\!=\!10$).  Between the bar and the
 spoke, which is visible in both the stars and the gas, there is
 a clear deficiency of gas.

 In both models a second expanding density wave ring becomes visible
 $\Delta t\!=\!20$ after the impact, when the bar has almost
 recentered.  The second ring in model \Model{B}{1} merges with a
 spiral arm segment, which has formed after the impact
 (Fig.~\ref{fig23}).
 In \Model{B}{2} the second ring is more symmetric than the first
 one and encloses the weak bar completely. However, the ring is
 not centered with respect to the bar. In fact the bar remains
 connected to the ring for a few dynamical times (Fig.\ref{fig24}).
 Similar to the previous simulations, additional ring-like structures
 or segments form in the gas at later times, merging with the spiral
 arms in the disc. These features are not supported by a stellar
 counterpart and are washed out relatively fast.

\section{Discussion}

\subsection{Bar amplitude}

 The evolution of the bar strength for the different models
 is shown in Fig.~\ref{fig19}~a) and b). In the isolated
 model the stellar bar forms within a few disc rotations and reaches
 its maximum strength at $t\!=\!60$.  As a result of the gas inflow
 the amplitude decreases rapidly at first, but then settles down to
 a quasi-stationary state, in which its strength decreases linearly
 on a time-scale which is significantly longer than the dynamical one,
 as described in \S3.1.

 The passage of the companion through the disc has several effects
 on the morphological evolution of the bar, such as a temporary 
 off-centering and a weakening or destruction of the bar. The 
 strength of these effects depends strongly on the impact location
 relative to the bar.  With the central passage the bar is destroyed
 very shortly after the impact of the companion ($\Delta t\!=\!10$,
 or 0.5 disc rotations).  For the early impact, the bar gets torn
 into two fragments which move outward with the expanding ring and
 later merge again, while for the late impact, no bar is left after
 the first stellar ring becomes visible. In both models the
 destruction of the bar is accompanied by a considerable vertical
 thickening of the disc.
 Impacts on the bar major axis (models of type A) show a similar
 evolution of bar strength as in the isolated model, although the
 bar is temporarily off-centered after the impact for some 0.6 Gyr.
 Comparison with the isolated model shows that the bar strength
 in these models is more significantly affected by the gas inflow
 than by the interaction itself. When the companion is close to
 the $z\!=\!0$ disc plane, the tidal force it exerts on the bar
 is directed almost parallel to the bar major axis and thus does
 not disrupt much the bar structure.
 In the models with the minor axis passage (models \Model{B}{1}
 and \Model{B}{2}) the bar is nearly destroyed, but its amplitude
 decreases more slowly ($\Delta t\!=\!50$) than in the corresponding
 central passage, an effect which is particularly evident in model
 \Model{B}{1}. The decay of the bar amplitude in both models can be
 described as quasi-exponential.  The dissolution of the bar is not
 accompanied by a vertical thickening of the disc in these models.
 For the minor axis impacts it seems to be possible to destroy
 the stellar bar, while keeping the disc, in contrast to the pure
 N-body simulations, where it is found that an interaction 
 sufficiently strong to destroy the bar also destroys the disc
 \cite{ath96b}, i.e. the ratio between the vertical and the radial
 scale-heights increased very significantly. This could be due to
 the fact that the decrease in the bar amplitude in the purely
 stellar cases can be due only to the impact, while in simulations
 including a gaseous component
 it is due to the cummulative effects of the impact and the gaseous
 central mass concentration.

 Central impacts and passages through the bar minor axis seem
 to cause more damage to the bar than the major axis impacts.
 The disc thickens more if it is hit at the centre than at the
 periphery so that the disc of models of type C thickens 
 considerably, while models of type A and B do not show a
 significant difference in this regard.  Between early and late
 impact models with same impact positions we find differences in
 the bar evolution shortly after the impact because of the
 different initial state of the bar at impact time, but the long
 term evolution of the bar in both types of models is similar.
 We conclude that the fate of the bar is more sensitive to the
 impact location of the companion relative to the bar than to the
 evolutionary phase of the bar during impact.
 In general the impact of a companion tends to accelerate the
 evolution from a strongly barred spiral to a weakly or non-barred
 galaxy, a trend also found in the isolated model, but on longer
 time-scales.

\subsection{Bar pattern speed}

 The evolution of the pattern speed in the isolated model is mainly
 influenced by two processes.  Angular momentum is transferred to
 the outer disc and halo resulting in a decrease of the pattern
 speed (e.g., Tremaine \& Weinberg 1984; Weinberg 1985;  Little \&
 Carlberg 1991; Hernquist \& Weinberg 1992; Athanassoula 1996b, 2002;
 Debattista \& Sellwood 1998, 2000), while
 mass accretion into the nuclear region steepens the rotation curve
 at the centre and leads to an increase of $\Omega_p$ (e.g., Friedli
 \& Benz 1993; Heller \& Shlosman 1994; Berentzen et al. 1998).
 In our isolated model we find that during the early phases of the
 bar evolution, i.e. during its growth and dissolution before  it
 has reached quasi-stability, the pattern speed increases linearly
 from $\Omega_{\rm p}\!=\!0.22$ at $t\!=\!30$ to
 $\Omega_{\rm p}\!=\!0.30$ at $t\!=\!80$,
 and thereafter stays constant to the end of the run. This behavior
 is in agreement with other numerical simulations 
 which have included a gas component (e.g., Friedli \& Benz 1993;
 Heller \& Shlosman 1994; Berentzen et al. 1998), showing either
 constant or slightly increasing pattern speeds.

 Due to the interaction we find for models with impacts on either the
 major or the minor axis of the bar some variations in the pattern
 speed (see Fig.~\ref{fig20}). In the late impact models, shortly
 after the impact, the
 pattern speed decreases temporarily by about $30\%$, but when the
 bar recenters again $\Omega_p$ increases back to its original
 value, though with oscillations that are damped over time.
 One of our models, however, shows an evolution in pattern speed
 which is different from that of our other vertical impact models.
 In model \Model{B}{1}, where the companion hits the disc on the
 minor axis of the strong bar, we find an increase of $\Omega_p$
 from 0.3$\tau^{-1}$ to 0.5$\tau^{-1}$.
 The bar maintains this high rotation after recentering and so
 has gained angular momentum. This effect is likely due to torques
 exerted by the perturber galaxy on the bar.
 For the central passage models \Model{C}{1} and \Model{C}{2} the
 bar seems to be slowed down by the interaction, although the
 $m\!=\!2$ amplitude of the bar is only marginal and a pattern speed
 is difficult to measure with confidence.
 The collisionless models studied by Athanassoula \shortcite{ath96a}
 and APB97 showed a decrease of the bar pattern speed in those
 models where an abrupt change of $\Omega_p$ occurred after the
 impact.  Sundin et al.  \shortcite{sds} found that depending on the
 mass of the perturber, both an increase or a decrease of
 $\Omega_p$ was possible. In their models, however, the trajectory
 of the perturbing galaxy lies in the equatorial plane of the disc.

\subsection{Off-centred bars}

 Another important feature produced by the interaction is the
 displacement of the bar from the centre of the galaxy.
 Asymmetries of this type are frequently found in disc galaxies,
 particularly in late type galaxies (e.g., de Vaucouleurs \& Freeman 1972;
 Odewahn 1996). While this kind of asymmetry could be produced by
 strong $m\!=\!1$ modes in the disc, an off-centred impact of a
 companion galaxy should generally give rise to more significant
 shifts, depending on the impact parameters.  In all our models
 with off-set passages we notice a strong displacement of the bar
 structure for approximately 0.6\,Gyr, independent of the evolutionary
 phase of the bar at impact time.  The maximum displacement of the bar
 centre, which is in our models roughly $\Delta R\!=\!3$\,kpc, depends
 only on the impact position or distance of the companion from the bar
 centre, but not on the strength of the bar at impact time. 
 The size of the displacement in our models is in good agreement with
 the relative displacement parameter found by Feitzinger \shortcite{fei80}
  for Magellanic type galaxies.
 The off-centring of the bar is also accompanied by asymmetries
 in the spiral structure. The morphology of the discs with
 off-centred bars in our simulations is dominated by one long
 spiral arm, which formed after the impact by an expanding density
 wave, in agreement with the results of Athanassoula \shortcite{ath96b}
 and with the morphology of several asymmetric barred galaxies as e.g.
 NGC\,4027.
 The induced asymmetries are visible in the disc, for about 0.6 Gyr
 after the impact. 

\subsection{Nuclear and circumnuclear disc}

 The presence of the bar results in a considerable change in the mass
 distribution of the gas. Due to the bar torques, gas is driven
 towards the centre of the galaxy and accumulates in nuclear
 and circumnuclear discs. These type of discs form in both the
 isolated and the interacting models, although their shape and size
 varies depending on the interaction parameters (Fig.~\ref{fig04}).
 The circumnuclear disc is always elongated perpendicular to the
 major axis of the bar, indicating that the gas in it populates
 the $x_2$ orbits and is trapped therefore between the inner and
 the outer ILR, which have been confirmed to be present by the
 surfaces of section.  The gas in the nuclear disc represents some
 18\% of the total dynamical mass inside the central region, i.e.
 within a radius of 1\,kpc.  This is about the same fraction as found
 in the isolated barred galaxy model by Berentzen et al. (1998).

 The cn-discs in the interacting models with central impact are much
 larger than in the isolated model (up to a factor of three in radius).
 Since the bar is destroyed by the interaction almost immediately in
 these models, there acts no torque on the gas to drive it closer to
 the centre.  In the major axis impacts the bar keeps roughly the same
 strength as in the isolated model and therefore we do not find a
 significant difference in the morphology of the central region.  Very
 little additional gas is driven towards the centre by the impact itself.
 
\subsection{Rings and spokes}

 In our central impact models \Model{C}{1} and \Model{C}{2} we find
 expanding ring structures in the disc, following the passage of
 the companion.
 The process of ring formation by interactions is described for
 non-barred galaxies in detail by Lynds \& Toomre \shortcite{lyt76}.
 These authors show that the expanding stellar ring is a density wave
 feature travelling through the disc and due to radial oscillations of
 the disc particles, as later confirmed by simulations (e.g., APB97).
 We find such radial
 oscillations for both the stellar and the gas particles,
 of which the latter contribute longer to the rings.  The gaseous ring
 always remains within the expanding stellar ring, having roughly the
 same radius, but considerably smaller width.
 The gaseous rings fragment on the inner side due to the
 self-gravity of the gas and its dissipative nature, while gas piles
 up in the ring during its expansion, because the gaseous
 orbits cannot cross. In contrast, the stellar ring becomes weaker and
 broader while expanding, but the gas ring can be identified for a longer
 time.

 The expansion velocity of the rings decreases with radius as
 predicted by the impulse approximation and as found in $N$-body
 simulations (APB97). The expansion velocity of the first ring
 drops gradually from 88 km\,s$^{-1}$ (7.5\,$v_{\rm s}$) to 35
 km\,s$^{-1}$ (3\,$v_{\rm s}$, where $v_{\rm s}$ is the sound
 speed in the gas). With the passage of the ring the local velocity
 dispersion in the disc increases.
 The second ring in our models starts off slower than the first, with 
 an expansion velocity which decreases from 29 km\,s$^{-1}$ 
 (2.5\,$v_{\rm s}$) to 11.72 km\,s$^{-1}$
 (1\,$v_{\rm s}$).  Note that there is a transition in the gas from
 a supersonic density wave supported by the stellar rings to an
 {\em acoustic} wave.

 A well known example of a collisionally induced ring galaxy is the
 Cartwheel galaxy (A0035-324).  Density enhancements like the rings
 are accompanied by the formation of bright young blue stars and
 H{\rm II} regions, while the induced stellar ring represents mainly
 the underlying disc population. The star-formation rate found in the
 outer ring of the Cartwheel galaxy is about $6.7 \cdot 10^7\,M_{\odot}
 {\rm Myr}^{-1}$ (Higdon 1995).
 After the passage of such star-forming rings, color gradients are
 expected in the disc and have been observed in several ring galaxies
 (Korchagin, Vorobyov \& Mayya 1999).
 In particular, such radial age gradients have been found in the
 optical and the NIR in the Cartwheel galaxy (Marcum, Appleton \&
 Hidgon 1992).  In numerical simulations which have included star
 formation \cite{mih94}, however, such gradients remained for only
 a short period of time, before the nuclear starburst dominated the
 rings.
 In our model \Model{C}{2} we find that the second gas ring, which
 forms after the impact, consists mainly of particles which had
 already contributed to the first one.  Thus, if we assume that some
 fraction of this gas is used up by star formation, we might expect
 the inner ring to contain considerably less gas than the outer ones,
 as confirmed by the observations of Cartwheel.

\subsection{Role of the halo}

 To study the role of the halo response in our simulations, we rerun a
 simulation of model \Model{C}{2}, but with a fixed (frozen) halo, i.e.
 we represented the halo by an external force, unchanged through the
 simulation.
 We find that the morphological evolution of the disc in face-on
 projection is very similar to the case with a live halo, e.g., the
 number of rings formed after the impact, their shape and their expansion
 velocity are qualitatively reproduced.
 The main difference between the two models is found in the vertical
 structure of the disc following the impact. In the edge-on projection
 we find that in the frozen halo model both the gaseous and the stellar
 disc become vertically thicker after the impact and settle down 
 in a more irregular shape, i.e. not disc-like, in contrast to the live
 halo model. In the latter we find that the halo gets centrally less
 concentrated and dynamically hotter after the impact.
 We therefore conclude that the live halo stabilizes the disc by absorbing
 energy from the impact.


\section{Summary}

 We have performed fully self-consistent N-body/SPH simulations to study
 the interaction between an initially barred galaxy and a less
 massive spherical companion. The companion passes {\em almost}
 vertically through the disc of the host galaxy and leaves the host
 system following the impact. The mass of the companion has been
 chosen so as to be a {\em significant} perturbation to the disc.
 Two different impact times have been chosen, representing
 different characteristic phases of the isolated bar evolution.
 In the first set of simulations the impact is at an early phase
 of evolution when the bar is strongest, while in the second set
 an advanced phase has been chosen, i.e.  when the bar is weak and
 has settled down in a quasi-stable state. Beside the impact time,
 we also varied the position of the impact with respect to the bar
 in the disc. In the simplest case the companion hits the bar at its
 centre, while in other cases it hits the disc outside the bar at
 its major or minor axis.  Interactions at times before a stellar
 bar has formed or during its non-linear growth phase have not been
 performed here.

 The interactions produce characteristic features, which remain in
 the disc for only a few dynamical times. In general the gaseous
 features tend to persist longer than their stellar counterparts,
 though after about 1\,Gyr both of them are dissolved.
 The most prominent features produced by the interaction are the
 collisionally induced rings, which are well known from non-barred
 galaxies.
 The central impacts produce expanding stellar and gaseous rings,
 whose symmetry is disturbed in the case of the strong bar model.
 The stellar ring is a density wave produced by forced radial
 oscillations of the stellar orbits.  The basic properties of the
 stellar rings, such as density or expansion velocity, are well
 described, at least qualitativly, by the impulse approximation.
 The gaseous rings also show the characteristics of a density wave,
 which is supported by the potential of the stellar ring.
 We find that more than two rings can be formed in the gaseous
 disc by the impact, some of them not supported by stellar
 density enhancements. These rings are less pronounced than
 the first two rings and dissolve much faster.
 In contrast to central impacts the off-centered passages produce
 outward-expanding ring-like density waves not in the form of
 closed rings, but as long tidal spiral arms whose extent exceeds
 the inital disc cut-off radius. These tidal arms are present for
 some few dynamical times in our simulations, before they dissolve and 
 the material in the arms is redistributed in the disc.

 In the simulations with off-centered impacts the stellar bar gets
 temporarily displaced from the center of the galaxy for some 0.6
 Gyr before it recenters again.  The gas inside the bar is driven
 towards the centre, forming a nuclear disc, whose mass is
 determined by the amount of gas located in the bar region.
 Almost no additional gas is driven
 towards the centre of the galaxy due to the interaction, compared
 to the isolated model.
 In most of our simulations the stellar bar is destroyed after the
 impact of the companion, leaving behind a dense stellar core
 at the centre of the galaxy. In the central impact models the destruction
 of the bar is accompanied with a considerable thickening of the
 disc. In models with minor axis impacts, however, it seems to be
 possible to destroy the bar while keeping the disc.
 We also find cases (i.e., the major axis impact) in which the bar
 survives the interaction, resembling the weak bar of the isolated
 model.  We argue that the interaction with a companion can drive
 the transition from a strongly barred galaxy to a weakly barred
 galaxy on shorter time-scales than those found in isolated 
 models.
 The impact time, or the dynamical phase of the disc, does not play
 a significant role in determining the final morphology of the disc.
 The dense nuclear disc and the surrounding gap in the gas distribution
 are the only morphological imprints which survive the interaction,
 even after the bar has dissolved. Such features in a non-barred galaxy
 may indicate the former existence of a bar.
 Apart from that, looking at the final face-on morphology of the
 gaseous disc -- in which the differences are more clear than in the
 stellar -- the interacting and non-interacting models can hardly be
 separated.  This might be possible from models including star
 formation, e.g., from the resulting color/metallicity gradients.


\section{ACKNOWLEDGMENTS}

 We would like to thank Albert Bosma for interesting discussions,
 J.C. Lambert and C. Theis for their computer assistance and an
 anonymous referee for comments which helped to improve the
 presentation.
 I.B. acknowledges DFG grant Fr 325/48-1, /48-2 which supported
 this work and the GRAPE facilities at the Sternwarte G\"ottingen.
 C.H. acknowledges support from DFG grant Fr 325/39-1, /39-2.  
 E.A. would like to thank the region PACA, the IGRAP, the INSU/CNRS
 and the University of Aix-Marseille {\rm I} for funds to develop the
 GRAPE facilities used for part of the calculations in this paper. 


\end{document}